\documentclass[journal]{IEEEtran}
\usepackage{array}
\usepackage{graphicx}
\usepackage{epsfig}
\usepackage{cite}
\usepackage{epstopdf}
\usepackage{amsfonts,amsmath,amsthm,amssymb,balance}
\usepackage{multirow,balance}
\usepackage{bm}
\usepackage[usenames,dvipsnames]{color}
\usepackage{colortbl}
\usepackage{float}
\usepackage{subfigure}
\usepackage{algorithm}
\usepackage{algpseudocode}
\usepackage{amsmath}
\usepackage{lipsum}
\usepackage{amsmath,epsfig,amssymb,subfigure,bm,dsfont}
\usepackage{lipsum}

\usepackage{stfloats}
\usepackage{cite}
\usepackage{amsmath}
\usepackage{mathrsfs}
\usepackage{epsfig}
\usepackage{subfigure}
\usepackage{caption}
\usepackage{epsf}
\usepackage{graphicx}   % jpg photo
\usepackage{psfrag}
\usepackage{amsfonts}
\usepackage{graphicx}
\usepackage{psfrag}
\usepackage{color}
\usepackage{amssymb}
\usepackage{multirow}
\usepackage{booktabs}
\usepackage{algorithm}
\usepackage{algorithmicx}
\usepackage{algpseudocode}
\usepackage{tabularx}
\usepackage{exscale}  % large Integral sign
\usepackage{relsize}  % large Integral sign
\usepackage{verbatim} % multiline comment
\usepackage{CJK}
\usepackage{indentfirst}
\usepackage[flushleft]{threeparttable}

\usepackage{color} %颜色

\definecolor{mygray}{gray}{.9}

\graphicspath{{figures/}}
\newcolumntype{C}[1]{>{\PreserveBackslash\centering}p{#1}}
\newcolumntype{R}[1]{>{\PreserveBackslash\raggedleft}p{#1}}
\newcolumntype{L}[1]{>{\PreserveBackslash\raggedright}p{#1}}

\usepackage{array}
\usepackage{graphicx}
\usepackage{epsfig}
\usepackage{cite}
\usepackage{epstopdf}
\usepackage{amsfonts,amsmath,amsthm,amssymb,balance}
\usepackage{multirow,balance}
\usepackage{bm}
\usepackage[usenames,dvipsnames]{color}
\usepackage{colortbl}
\usepackage{float}
\usepackage{subfigure}
\usepackage{algorithm}
\usepackage{algpseudocode}

\ifCLASSINFOpdf
\else
\fi
\begin{document}
\title{A Joint Beamforming Design and Integrated CPM-LFM Signal for Dual-functional Radar-communication Systems}

\author{Yu~Cao, ~\IEEEmembership{Student~Member,~IEEE},
        Qi-Yue~Yu, ~\IEEEmembership{Senior~Member,~IEEE}

\thanks{
The work presented in this paper was supported by the National Natural Science Foundation of China under Grand No. 62071148, and partly supported by the Natural Science Foundation of Heilongjiang Province of China under Grand No. YQ2019F009.(Corresponding author: Qiyue Yu).
}

\thanks{
Y.~Cao and Q.-Y.~Yu are with the Communication Research Center, Harbin Institute of Technology, China. (email: \{yucao, yuqiyue\}@hit.edu.cn). }
}

\markboth{IEEE TRANSACTIONS ON SIGNAL PROCESSING,~Vol.~XX, No.~X, Aug.~2021}%
{Shell \MakeLowercase{\textit{et al.}}: Bare Demo of IEEEtran.cls for IEEE Journals}

\maketitle

\begin{abstract}
The dual-functional radar-communication (DFRC) system is an attractive technique, since it can support both wireless communications and radar by a unified hardware platform with real-time cooperation. Considering the appealing feature of multiple beams, this paper proposes a precoding scheme that simultaneously support multiuser transmission and target detection, with an integrated continuous phase modulation (CPM) and linear frequency modulation (LFM) signal, based on the designed dual mode framework. 
Similarly to the conception of communication rate, this paper defines radar rate to unify the DFRC system. 
Then, the maximum sum-rate that includes both the communication and radar rates is set to be the objective function. Regarding as the optimal issue is non-convex, the optimal problem is divided into two sub-issues, one is the user selection issue, and the other is the joint beamforming design and power allocation issue.
A successive maximum iteration (SMI) algorithm is presented for the former issue, which can balance the performances between the sum-rate and complexity; and maximum minimization Lagrange multiplier (MMLM) iteration algorithm is utilized to solve the latter optimal issue.
Moreover, we deduce the spectrum characteristic, bit error rate (BER) and ambiguity function (AF) for the proposed system. Simulation results show that our proposed system can provide appreciated sum-rate than the classical schemes, validating the efficiency of the proposed system.
\end{abstract}

\begin{IEEEkeywords}
Dual-functional radar-communication (DFRC), precoding, continuous phase modulation (CPM), linear frequency modulation (LFM), maximum minimization Lagrange multiplier (MMLM), user selection.
\end{IEEEkeywords}

\IEEEpeerreviewmaketitle

\section{Introduction}
\IEEEPARstart{C}{urrently}, the joint communication and radar (JCR) system has been widely researched in both academic and industry \cite{JCR-scenario}, since it can simultaneously satisfy the data transmission and target detection, providing real-time information sharing. Therefore, the JCR system has been applied into indoor positioning \cite{vehicles_1}, UAV networks \cite{UAV}, vehicle networks \cite{vehicles_2}, and etc. Especially, the JCR system is an appealing choice for seamless connectivity to serve multiple communication users and achieve the target information. Compared to the classical single radar or communication equipment, it has advantages in terms of cost, size, power consumption, spectrum usage, and etc \cite{CRI_advantages} \cite{CRI_overview}.

The concept of JCR was first proposed by Space Shuttle program of National Aeronautics and Space Administration (NASA) in the late 1970's \cite{CRI_present}. The initial goal was to integrate the electronic warfare (EW) of communication and radar functions into the antenna array (AA) for resource sharing, mainly used in military scenarios. Slowly, JCR is applied into civil area, e.g., advanced driver assistance systems (ADAS) \cite{Autonomous Driving}. In airborne areas, \cite{Airborne} merges the communication system and the synthetic aperture radar (SAR) by the space time coding (STC) scheme.

The prospective research directions of JCR mainly include radar-communication coexistence system (RCC) and dual-functional radar-communication (DFRC) \cite{JCR-survey}. The RCC system emphasizes on the coexistence of communication and radar system that mainly develops efficient interference management techniques, including null space signal projection, transmit waveform design, power-efficient pattern design and dynamic spectrum allocation \cite{RCC-4}-\!\!\cite{RCC-6}. \cite{RCC-4} proposes a null space projection method to eliminate the interferences of the RCC system, with the knowledge of the channel state information (CSI). \cite{RCC-5} designs transmit waveforms by space-time code based on the controllable degrees of freedom to reduce the self-interferences of the RCC receiver. For the signal side-lobe interference control aspect, \cite{Sidelobe Reduction} presents a Riemann gradient conjugate power budget waveform design method to reduce the influence of side-lobe interference of a multiple input multiple output (MIMO) RCC system. In \cite{RCC-6}, the authors aim to minimize the downlink multiuser interference by using branch-and-bound dynamic spectrum allocation method.

On the other hand, the DFRC focuses on the cooperation of communication and radar systems, and generally adopts integrated signal to improve the performances of both communication and radar \cite{JCR-survey-2}. There are many prospective research directions of DFRC, especially, beamforming design and integrated signal construction are two of the appealing ones \cite{DSRC-diversity}.

There are many works on beamforming design of the DFRC system \cite{DSRC_beamforming_1}-\!\!\cite{DSRC-beamforming-4}. In \cite{DSRC_beamforming_1}, the authors present an optimal beamforming method to minimize the downlink multiuser interference, and further propose a weighted beam-pattern for a flexible trade-off between the performance of radar and communication. In \cite{DSRC-beamforming-5}, the authors design a beamformer that matches the radar's beam-pattern; moreover, it satisfies the communication performance relied on the null-space criterion with the constraint of signal-to-interference-plus-noise ratio (SINR). \cite{DSRC-beamforming-2} presents a predictive beamforming scheme to enhance angle estimation accuracy by the maximum posteriori probability criterion, which can acquire the location and track the information of a vehicular system. In \cite{DSRC-beamforming-3}, the authors present a hybrid beamformer based on the minimum mean square error (MMSE) criterion, which can enhance the achievable sum-rate significantly. \cite{DSRC-beamforming-4} presents the transmit beamforming scheme based on the maximum SINR criterion, which optimizes both the radar transmit beam pattern and communication rate.

Besides the beamforming topic, signal waveforms designing is another hot topic. To combine the communication and radar waveforms, there are generally three major categories, they are orthogonal frequency division multiplexing (OFDM), frequency modulation (FM) and spread spectrum (SS) \cite{JCR-overview}. In \cite{mutual information}, the author proposes an OFDM queue scheduling model, taking both the network stability and radar detection performance into consideration. \cite{CD-OFDM} proposes a novel code-division OFDM system to provide high spectrum efficiency and robust radar sensing performances. The linear frequency modulation (LFM) is widely considered in DFRC systems, since it provides a larger detection range and a higher range resolution for radar detection \cite{DSRC-signal-2}. Phase-coded frequency modulation can mitigate the impact of Doppler shift, thus it is suitable for high speed target detection \cite{PCFM}. In \cite{High-Speed}, the authors combine the low density parity check (LDPC) codes with LFM signals for a DFRC system, which can provide a better BER than the LFM signal. The combination of continue phase modulation (CPM) and LFM can reduce the influence of side lobes \cite{DSRC-CPM}, thus improve the power and spectrum efficiency. As for spread spectrum signal, the direct sequence (DS) can provide well-behaved auto- and cross-correlation properties to separate the radar and communication information perfectly \cite{SS-1}.

Obviously, a joint design of beamforming and signal-form for a DFRC system is a prospective research topic, since it can take signal processing gains to avoid interferences and improve the spectrum efficiency. \cite{beamforming-signal-1} considers the Hadamard-Walsh orthogonal codes with minimum variance distortionless response (MVDR) beamforming algorithm to achieve a better jamming and interference mitigation capability. \cite{beamforming-signal-2} presents a zero-forcing beamforming algorithm to transmit the independent radar waveforms and communication symbols simultaneously.

Considering the attractive feature of the joint design, this paper presents a joint beamforming and integrated CPM-LFM signal for a DFRC system. 
The major advantages of CPM are two aspects: one is the continuity of signal phase that improves the spectrum; and the other is that the Viterbi decoding (or demodulation) provides better BER performance. Furthermore, LFM signal is widely used in radar detection, since it has large time-bandwidth product, which can effectively solve the contradiction between resolution and measuring accuracy. 
Thus, the integrated CPM-LFM signal is an appealing attempt for a JCR system. 
To further improve the performance of CPM-LFM, antenna array is jointly considered in this paper, and the precoder is designed based on the maximum sum-rate that includes both communication and radar rates as our objective function. The contributions of this paper are mainly summarized as three aspects:
\begin{enumerate}
\item We present a joint beamforming design and integrated CPM-LFM signal for JCR systems. The designed dual mode framework that is consisted of static and dynamic beams can simultaneously support multiuser transmission and target detection;
\item Based on the definition of radar rate, we set maximum sum-rate as our objective function that can be proved to be a non-convex optimal issue. To find the optimal solution, the entire problem is divided into two sub-issues, which are user selection and beamforming weights design with power allocation. A successive maximum iteration algorithm is proposed for the user selection, and the maximum minimization Lagrange multiplier (MMLM) is proposed to solve the latter one; and
\item Theoretical analysis are presented for the proposed system, including spectrum, bit error rate (BER) and ambiguity function (AF) analysis. Numerical results verify the validity of the proposed system.
\end{enumerate}

The remainder of this paper is organized as follows. Section II presents the system model and the dual mode framework of the proposed system. In Section III, the transmitter and receiver of the proposed system is described in details. An objective function and its solutions are deduced in Section IV. Section V discusses the spectrum characteristic, BER and AF. Simulation results are given in Section VI, following with some conclusion remarks in Section VII.

\section{Preliminary}\label{section2}
In this paper, denote ${\mathbb{B}}$ and ${\mathbb{C}}$ by binary field and complex field. Let $x$, $\bf{x}$ and $\bf{X}$ be a scalar, a vector and a matrix, respectively. The transposition, conjugate and conjugate transpose of $\bf{x}$ are donated by $\bf{x}^{\rm{T}} $, $\bf{x}^*$ and $\bf{x}^{\rm{H}} $ respectively. Define $*$ as convolution operation, and $\Vert {\bf{x}} \Vert _2$ represents vector Euclidean norm. $\mathcal{CN}(\mu,\sigma^{2})$ indicates the complex Gaussian distribution with mean $\mu$ and covariance $\sigma^{2}$. $ {\rm Re}(x) $ means the real part of the complex number $x$. ${\mathbb{E}(\bf{x})}$ means the expectation of a random sequence $\bf{x}$. ${\bf I}_n$ is $n$-dimensional identity matrix, $\rm{Det}(\bf{X})$ represents the determinant of $\bf{X}$, $\rm{tr}(\bf{X}) $ means the a trace of $\bf{X}$, $\rm{Diag}(\bf{X})$ means a diagonal matrix that is composed of the diagonal element of $\bf{X}$, and $\rm{vec}(\bf{X})$ indicates converting a matrix $\bf{X}$ to a vector by vectorizing columns from the left side to the right side of $\bf{X}$.

\subsection{System model}
Assume the base station (BS) is equipped with $N_t$ transmit antennas and $N_r$ received antennas, which form transmit and received antenna arrays for both communication and radar functions. There are totally $U$ user equipments (UEs), where $U \gg N_t$, and each UE holds a single antenna. Since the BS cannot serve all the UEs simultaneously, we assume that $K$ UEs are selected from $U$, where $K \le N_t$. 
The locations of all the $U$ UEs are assumed to be available at the BS. Let the range, speed, elevation and azimuth angle of the target (TA) be $d_T$, $v_T$ and $(\theta_T, \varphi_T)$ respectively, which will be estimated by the radar at the BS. The ranges of elevation $\theta_T$ and azimuth $\varphi_T$ are respectively $(0, \pi/2)$ and $(0, 2\pi)$.

\begin{figure*}[th]
\centering
\includegraphics[width = 6.2 in]{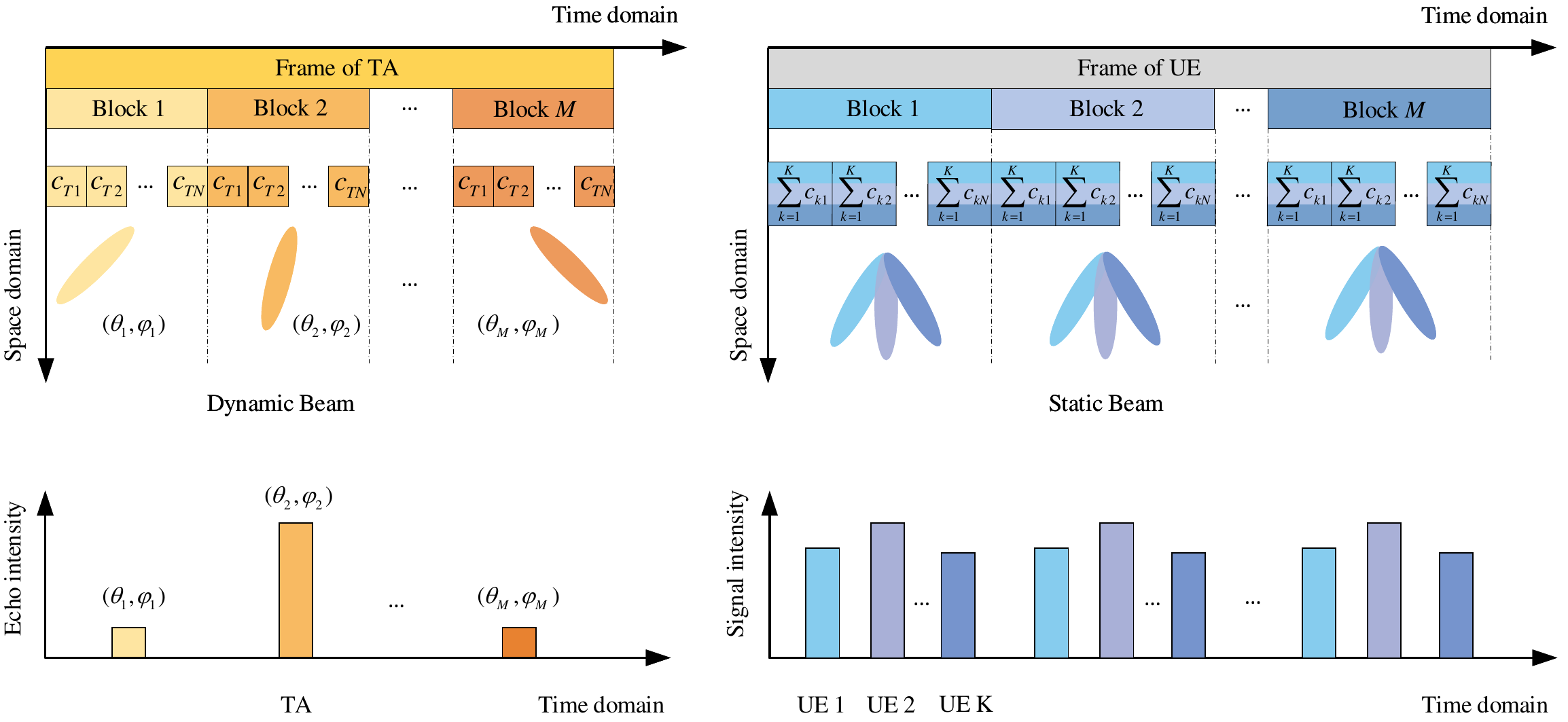}
\caption {A dual mode framework of a DFRC system, including time-sharing scanning mode given in (a) and fixed-direction mode given in (b). }
\label{Frameworks}
\end{figure*}

The transmit antennas of the BS is set to be a uniform rectangular array (URA) \cite{URA}, whose  element owns a separate phase shifter, providing a three-dimensional (3D) beamforming from elevation and azimuth dimensions. Suppose the numbers of the elements of the transmit URA placed along the X-axis and Y-axis are respectively $N_{tx}$ and $N_{ty}$, indicating $N_t = N_{tx} \times N_{ty}$. Let the distances between adjacent antenna elements of X-axis and Y-axis be $d_x$ and $d_y$ respectively. Denote the label of each transmit antenna element by $(n_{tx},n_{ty})$, where $1\le n_{tx}\le N_{tx}$ and $1\le n_{ty}\le N_{ty}$. The label of the $n_t$th element can be expressed as ${n_t} = {(n_{ty}-1)} {N_{tx}} + {n_{tx}}$, where $1\le n_t \le N_t$.

Similarly, the received antenna array of the BS is also set to be a URA, and the elements of the received URA are respectively $N_{rx}$ and $N_{ry}$, i.e., $N_r = N_{rx} \times N_{ry}$. The label of the $n_r$th received antenna is ${n_r} = {(n_{ry}-1)} {N_{rx}} + {n_{rx}}$, where $1\le n_{rx}\le N_{rx}$, $1\le n_{ry}\le N_{ry}$ and $1\le n_r \le N_r$.

\subsection{Framework}
The transmit URA of the BS should simultaneously provide different beams for both UEs' communication and TA's detection. The UEs' angle information are available at the BS, and the beamforming algorithms are determined by these angle information. On contrast, the TA's angle information is unknown, which needs to be estimated. Thus, the transmit URA of the BS is consist of two types of beams, one is dynamic beams for TA's searching, and the other is static beams for UEs' communications, as shown in Fig. \ref{Frameworks}.

Through dynamic beam scanning, both the position and speed information of the TA are expected to be estimated. Meanwhile, UEs are served by the static beams. Because of the different roles of dynamic and static beams, this paper proposes a dual mode framework that is consisted of dynamic time-sharing scanning mode and static fixed-direction mode.

Assume one frame is consisted of $M$ blocks and each block includes $N$ symbols. It is noted that $M$ is equal to the product of $M_e$ and $M_a$, i.e., $M_e = \frac{\pi}{2 \delta_\theta}$ and $M_a = \frac{2\pi}{\delta_\varphi}$, where ${\delta_\theta}$ and ${\delta_\varphi}$ are beamwidth in elevation and azimuth dimensions. Therefore, the TA detection precision relays on the value of $M$. A large $M$ indicates a higher precision, at the cost of longer scanning time. The dynamic beam alters its direction every block time, thus named as ``dynamic". During one frame time, there are totally $M$ dynamic beams. Conversely, the static beams keep the directions during the entire frame time, as if ``static". Dynamic beam utilize time-sharing scanning mode to detect the position and speed of a moving TA. The scanning direction $(\theta_m, \varphi_m)$ of the $m$th block, where $1 \le m \le M$, keeps as a constant during the $m$th block. If the reflected echo signal of the $m$th dynamic beam that is larger than the threshold, the TA is estimated to be at the direction $(\theta_m, \varphi_m)$. If the reflected echo signal of the $m$th dynamic beam is smaller than the threshold, there is no TA exists.

% section-3
\section{Design of joint beamforming and CPM-LFM integrated waveform}\label{section3}
\begin{figure*}[t]
\centering
\includegraphics[width = 7.1 in]{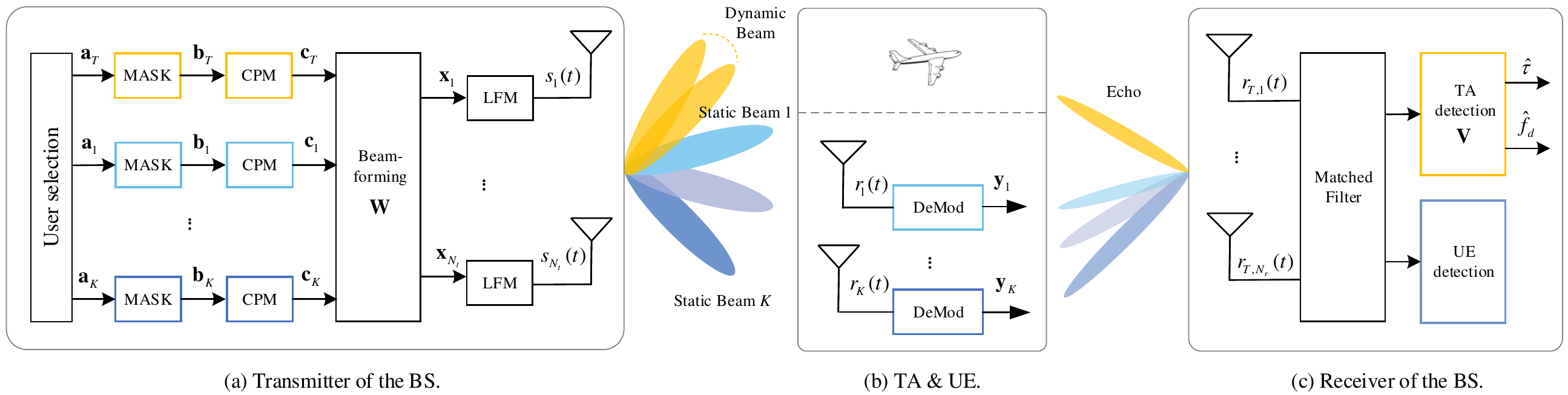}
\caption {A diagram of the transmitter and receiver of our proposed DFRC system, where ``DeMod" indicates demodulation. There are $N_t$ transmit antennas, and $N_r$ received antennas at the BS. Each UE is equipped with one antenna. }
\label{System Model}
\end{figure*}

This section presents a joint beamforming and CPM-LFM integrated system that is designed based on the dual mode framework, as shown in Fig. \ref{System Model}. 
% Here is Figure-3 System Model

\subsection{Transmit signals of the BS}
To detect whether there is a TA or not, a pilot sequence is transmitted by the $m$th dynamic beam, and defined by $ {\bf{a}}_T = \left[ a_{T,1}, a_{T,2},...,a_{T,N {\rm log} \cal{M} } \right] \in \mathbb{B}^{1 \times N {\rm log}\cal{M} }$, where $\cal{M}$ is modulation order of a $M$-ary amplitude shift keying (MASK), and the subscript ``T" indicates the sequence (or signal) for the TA detection. Suppose $K$ selected UEs are simultaneously served by the static beams, where $1 \le K \le U$, define the transmit binary data sequence of the $k$th UE by $ {\bf{a}}_k = \left[ a_{k,1}, a_{k,2},...,a_{k, N {\rm log}\cal{M} } \right] \in \mathbb{B}^{1 \times N {\rm log}\cal{M}}$, where $1 \le k \le K$. The binary data sequences are firstly passed to the MASK mapper, and achieved complex signals $ {\bf{b}}_T = \left[ b_{T,1}, b_{T,2},...,b_{T,N} \right]\in \mathbb{C}^{1 \times N}$ and $ {\bf{b}}_k = \left[ b_{k,1}, b_{k,2},...,b_{k,N} \right]\in \mathbb{C}^{1 \times N}$.

Then ${\bf b}_T$ and ${\bf b}_k$ are passed to the CPM modulator. In order to ensure the phase continuity, the phase of the baseband signal of TA and the $k$th UE at the $n$th symbol duration $T_s$ are calculated as 
\begin{equation}
\begin{split}
    \beta_{T,n} = \sum_{i=1}^{n} b_{T,i} h \pi,\\
    \beta_{k,n} = \sum_{i=1}^{n} b_{k,i} h \pi,
\end{split}
\end{equation}
where $1 \le n \le N$ and the parameter $h$ is modulation index \cite{modulation_index}. The corresponding phase vectors of the TA and the $k$th UE are $\boldsymbol{\beta}_T = [\beta_{T,1}, \beta_{T,2},...,\beta_{T,N}]$ and $\boldsymbol{\beta}_k = [\beta_{k,1}, \beta_{k,2},...,\beta_{k,N}]$, respectively. Thus, the baseband signals of CPM are obtained as $ {\bf{c}}_T = e^{ j {\boldsymbol{\beta}}_T} \in \mathbb{C}^{1 \times N}$ and $ {\bf{c}}_k = e^{ j {\boldsymbol{\beta}}_k} \in \mathbb{C}^{1 \times N}$, forming the transmit data matrix ${\bf{C}} = \left[ {\bf c}_T^{\rm T},{\bf c}_1^{\rm T},...{\bf c}_K^{\rm T} \right]^{\rm T} \in \mathbb{C}^{(K+1)\times N} $.

In this paper, the transmit URA adopts fully connection structure. Suppose that the beamforming weight vectors of the TA and the $k$th UE are defined by ${\bf{w}}_T = [w_{1,T},...,w_{n_t,T},...,w_{N_t,T}] \in \mathbb{C}^{N_t \times 1}$ and ${\bf{w}}_k = [w_{1,k},...,w_{n_t,k},...,w_{N_t,k}] \in \mathbb{C}^{N_t \times 1}$, thus the precoding matrix can be represented as $ {\bf{W}} = [{\bf{w}}_T, {\bf{w}}_1,...,{\bf{w}}_K ] \in \mathbb{C}^{N_t \times (K+1)}$. Assume the transmit power of the TA and $k$th UE are $P_T$ and $P_k$ respectively, and the corresponding power allocation matrix is $ {\bf P} = {\rm Diag} \left(P_T,P_1,...,P_K \right) \in \mathbb{C}^{(K+1) \times (K+1)} $. After precoding and power allocation, the transmit data on the $n_t$th antenna is represented as
\begin{equation}\label{x_n_t}
  {\bf x}_{n_t} = \sqrt{P_T} \, w_{n_t,T} \, {\bf c}_T + \sum_{k=1}^{K} \sqrt{P_k} \, w_{n_t,k} \, {\bf c}_k,
\end{equation}
where ${\bf x}_{n_t} = [x_{n_t,1},...,x_{n_t,n},...,x_{n_t,N}] \in \mathbb{C}^{1 \times N}$ and $1 \le n_t \le N_t$. 
% For the sake of simplicity, we choose rectangular pulse with amplitude $\frac{1}{2 T_s}$ to satisfy the condition of unit symbol energy, which can be achieved as
% \begin{equation}
%   g(t)=
%   \begin{cases}
%       \dfrac{1}{2 T_s}, & 0 \le t \le T_s, \\
%       0, & \text{others}.
%   \end{cases}
% \end{equation}
The baseband signal ${\bf x}_{n_t}$ is passed to the filter, and modulated to the carrier frequency $f_c$ to form LFM waveforms, then the corresponding integrated bandpass signal is ${\bf s}(t) = \left[{s_1}(t),...,{s_{n_t}}(t),...,s_{N_t}(t) \right]^{\rm T}$, where $s_{n_t}(t)$ is transmitted on the $n_t$th antenna, which is given by
\begin{align}\label{CPM-BF-LFM}
s_{n_t}(t) = \sum_{n=1}^{N} x_{n_t,n} \, & {g \left( t - (n-1) T_s\right)} \, {\rm e}^{ j \left(2\pi f_c t + \pi \mu t^2\right) },
\end{align}
where $(n-1)T_s \le t \le n T_s$. $g(t)$ is the transmit pulse, which can be rectangular pulse, raised cosine pulse, Gaussian pulse, and etc \cite{pulse}. $\mu$ is the chirp rate of the bandpass integrated signal \cite{chirp_rate}. If $\mu > 0$, it is a up-chirp signal; otherwise it is a down-chirp signal.

\begin{figure}[t]
\centering
\includegraphics[width = 3.5 in]{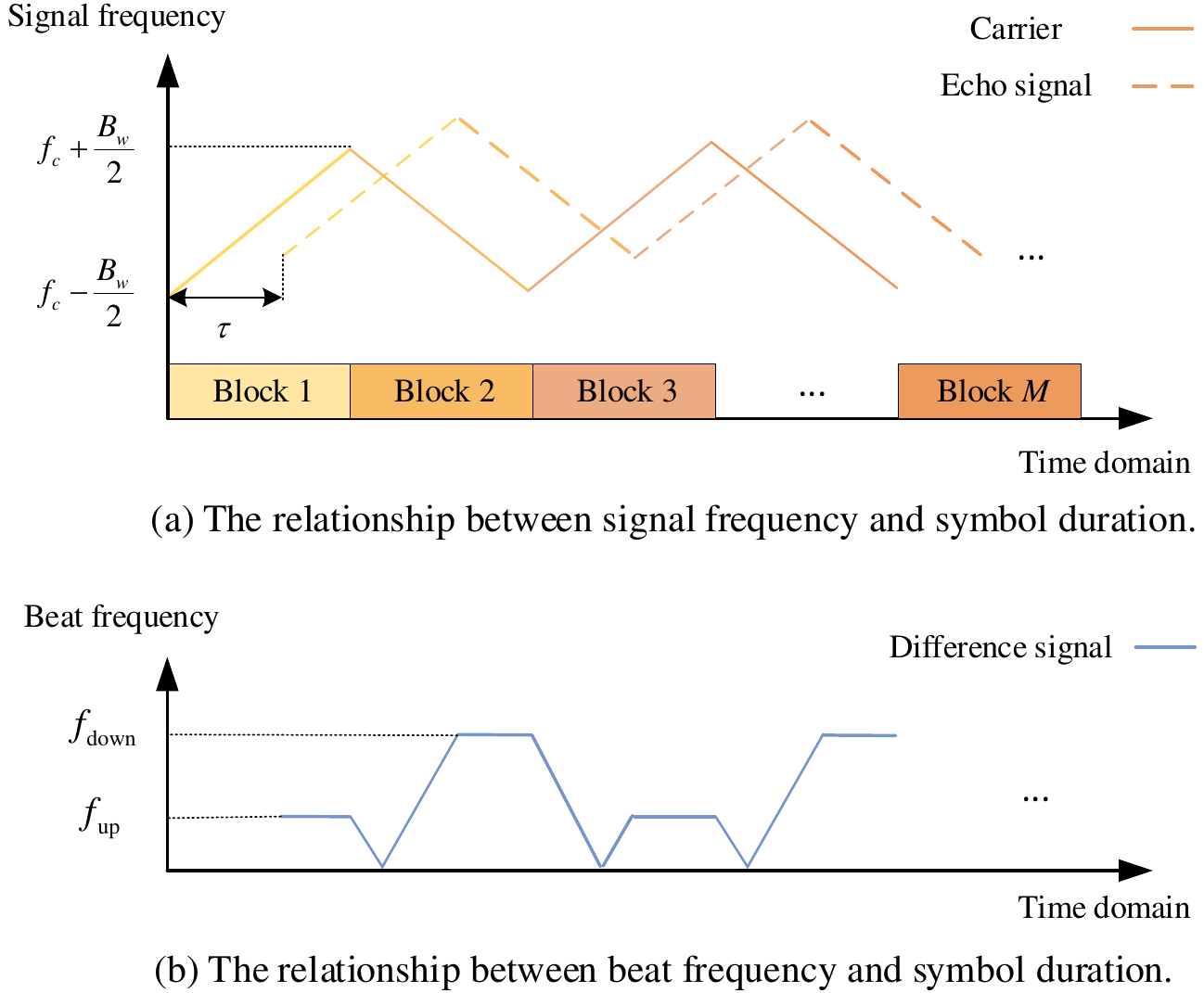}
\caption {The frequency variations of the carrier, echo signal and difference signal during $M$ blocks, where $\tau$ is the latency, $f_c$ is the center frequency and $B_w$ is the sweep bandwidth. }
\label{radar detection}
\end{figure}

Due to the feature of the triangular wave, the chirp rate is a constant value during one block time $T_B$, i.e., $T_B = N T_s$, and the chirp rate changes to its opposite value during the next block time. As shown in Fig. \ref{radar detection} (a), there are $M$ blocks, and the chirp rates of the odd and even blocks are respectively $\mu$ and $-\mu$, which are also called up-chirp and down-chirp blocks. 
% Here is Figure-5 radar detection

According to the chirp rate $\mu$ and block time $T_B$, the sweep bandwidth $B_w$ is given by
\begin{equation}\label{bandwidth}
    B_w = \mu \, T_{B} = \mu N T_s, 
\end{equation}
which indicates the highest frequency and lowest frequency of the transmit signals are respectively $f_c  + B_w/2$ and $f_c - B_w/2$.

\subsection{Received signals at each UE}
The integrated signal is then transmitted to the fading channel. Assume that the elevation and azimuth of the $k$th UE $\theta_k$ and $\varphi_k$ are perfectly known at the BS, and the ranges of elevation $\theta_k$ and azimuth $\varphi_k$ are respectively $(\pi/2, \pi)$ and $(0, 2\pi)$. Thus, the CSI ${\bf{h}}_{k} \in \mathbb{C}^{1 \times N_t}$ between the $k$th UE and BS is modeled as
\begin{equation}
    {\bf{h}}_{k} = \sqrt{ \left(d_k \big/ d_0 \right)^{-\alpha} \delta_k} \, {\bf{a}}_t(\theta_k,\varphi_k),
\end{equation}
where $d_k$ is the distance from the BS to the $k$th UE, $d_0$ is the reference distance, $d_k \geq d_0$, $\alpha$ is the path loss exponent, $\delta_k$ indicates shadowing effect that is a zero mean and $\sigma_\delta$ variance log-normal random variable. $ {\bf{a}}_t(\theta_k,\varphi_k) = \mathrm{vec} \left( { {\bf{a}}_{tx}(\theta_k,\varphi_k)}^{\rm{T}} {\bf{a}}_{ty}(\theta_k,\varphi_k) \right) \in \mathbb{C}^{1 \times N_t} $ is the transmit steering vector from the BS to the $k$th UE \cite{steering_vector}, which are given by (\ref{steering_vector}), where $\lambda = \frac{c}{f_c} $ stands for the wavelength.
\begin{figure*}[t]
\setcounter{equation}{5}
\begin{equation}\label{steering_vector}
\begin{split}
{\bf{a}}_{tx}(\theta_k,\varphi_k) = \frac{1} { \sqrt{N_{tx}} } \left[1, e^{j 2\pi \frac{d_x}{\lambda} \cos{\theta_k}\cos{\varphi_k}},...,e^{j 2\pi ({N_{tx}-1}) \frac{d_x}{\lambda} \cos{\theta_k}\cos{\varphi_k}} \right], \\
{\bf{a}}_{ty}(\theta_k,\varphi_k) = \frac{1} { \sqrt{N_{ty}}} \left[1, e^{j 2\pi \frac{d_y}{\lambda} \cos{\theta_k}\sin{\varphi_k}},...,e^{j 2\pi ({N_{ty}-1}) \frac{d_y}{\lambda} \cos{\theta_k}\sin{\varphi_k}} \right].
\end{split}
\end{equation}
\hrulefill
\end{figure*}

At the receiver of the $k$th UE, the received bandpass signal $r_k(t)$ can be given as
\begin{equation}\label{eq1}
     r_k(t) =  {\bf h}_k \, {\bf s}(t) + n_k(t).
\end{equation}

The received bandpass signal $r_k(t)$ is coherent demodulated and passed through the lowpass filter, the equivalent baseband signal $ {\bf y}_k \in {\mathbb C}^{1 \times N}$ is
\begin{align}\label{comm_signal}
    {\bf{y}}_k &= \sqrt{P_k} \, {\bf h}_k {\bf w}_k {\bf c}_k + \sqrt{P_T} \, {\bf h}_k {\bf w}_T {\bf c}_T \notag \\ & + \sum_{i=1,i\neq k}^{K} \sqrt{P_i} \, {\bf h}_k {\bf w}_i {\bf c}_i + {\bf n}_k,
\end{align}
where ${\bf{n}}_k$ is the complex additive white Gaussian noise (AWGN) obeying $\mathcal{CN}(0,\sigma_{k}^{2})$ with variance $\sigma_k^2$. The first item is our expected signal, the second and the third items are both interferences, which can be eliminated by designing precoding matrix $\bf X$, and the last one is noise.

\subsection{Received signals at the BS}
If there is a TA at the location of $(\theta_m, \varphi_m)$ during the $m$th block time, an echo signal is feedback to the BS. The received URA of the BS is used to acquire the echo signals of the TA. The steering vectors from the BS to the TA is denoted by ${\bf{a}}_t(\theta_m,\varphi_m) = \mathrm{vec} \left( { {\bf{a}}_{tx}^{\rm{T}}(\theta_m,\varphi_m)} {\bf{a}}_{ty}(\theta_m,\varphi_m) \right) \in \mathbb{C}^{1 \times N_t} $, and from the TA to BS is ${\bf{a}}_r(\theta_m,\varphi_m) = \mathrm{vec} \left( { {\bf{a}}_{rx}^{\rm{T}}(\theta_m,\varphi_m)} {\bf{a}}_{ry}(\theta_m,\varphi_m) \right) \in \mathbb{C}^{1 \times N_r}$. Let ${\bf A} = {\bf a}_r^{\rm T}(\theta_m,\varphi_m) {\bf a}_t(\theta_m,\varphi_m) \in {\mathbb C}^{N_r \times N_t} $, and ${\bf A}_K = \left[{\bf{a}}_r^{\rm T}(\theta_1,\varphi_1), {\bf{a}}_r^{\rm T}(\theta_2,\varphi_2),...,{\bf{a}}_r^{\rm T}(\theta_K,\varphi_K) \right] \in {\mathbb C}^{N_r \times K}$, where ${\bf{a}}_r^{\rm T}(\theta_k,\varphi_k)$ is the steering vector from the $k$th UE to the BS.

Thus the received bandpass signal at the BS is
\begin{equation}
    {\bf r}(t) =  {\cal L} {\bf A} \, {\bf s}(t-\tau) {\rm e}^{j 2 \pi f_d t} + {\bf A}_K {\bf I}_K(t) + {\bf n}_T(t),
\end{equation}
where ${\bf r}(t) = \left[r_1(t),...,r_{n_r}(t),...,r_{N_r}(t) \right]^{\rm T}$, ${\cal L} = \sqrt{ {\cal L}_T {\eta_T}} $ is the gain of echo signal, ${\cal L}_T$ is the propagation attenuation and $\eta_T$ represents radar cross section (RCS). $\tau$ and $f_d$ are respectively latency and Doppler frequency of the TA, which are need to be estimated. Let ${\bf I}_K = \left[ I_1(t), I_2(t),..., I_K(t) \right]$, where $I_k(t)$ is the signal of the $k$th UE, which is viewed as interferences for the TA detection, the power of which satisfy $\mathbb{E} \left( \left| I_k(t) \right|^2 \right) = 1$. ${\bf n}_T(t)$ is the complex AWGN obey $\mathcal{CN}(0,\sigma^2)$ with variance $\sigma^2$.

Define the processing vector of the BS by ${\bf V} \in {\mathbb C}^{1 \times N_r}$, thus, we have
\begin{align}
    {\bf y}(t) &= {\bf V} \, {\bf r}(t), \notag \\
      &= {\cal L} {\bf V} {\bf A} {\bf W} {\bf C} \cdot {\rm e}^{j [2 \pi f_c (t-\tau) + \pi \mu (t-\tau)^2 + 2 \pi f_d t] } \\  &+ {\bf V} {\bf A}_K {\bf I}_K(t) + {\bf n}(t) \notag,
\end{align}
where ${\bf n}(t) = {\bf V} \, {\bf n}_T(t)$ is the equivalent noise after signal processing.

Define the received difference signal ${\bf d}_r(t)$ by the low-pass signal of the product of the echo signal ${\bf y}(t)$ and the coherence frequency carrier, i.e., ${\bf d}_r(t) = {\rm LPF} \big[{\bf y}(t) \cdot {\rm e}^{ j \left(2\pi f_c t + \pi \mu t^2\right) } \big]$, where ${\rm LPF}$ represents low-pass filter. The frequencies of ${\bf d}_r(t)$ are defined by beat frequencies, including $f_{\rm up}$ and $f_{\rm down}$, which are respectively calculated in the up-chirp and down-chip stages, as shown in Fig. \ref{radar detection} (b). In each up-chirp block, the frequency of the echo signal is lower than the carrier owing to the influence of latency and Doppler frequency, while in the down-chirp block, the frequency of the echo signal is higher than the carrier.

Thereby, during the up-chirp stage, the received difference signal ${\bf d}_{r,\rm up}(t)$ can be expressed as
\begin{align}\label{SINR_T}
  {\bf d}_{r,\rm up}(t) &= {\rm LPF} \left[{\rm Re} \left({\bf y}(t) \right) \cdot {\rm Re} ({\rm e}^{ j \left(2\pi f_c t + \pi \mu t^2\right) }) \right],  \notag \\
  &= {\cal L} {\bf V} {\bf A} {\bf W} {\bf C} \cdot {\rm e}^{j [2 \pi (\mu \tau- f_d) t + 2 \pi (f_c \tau + \frac{1}{2} \mu \tau) \tau) ]} \notag \\&+  {\bf V} {\bf A}_K {\bf I}_K(t) + {\bf n}(t),
\end{align}
where the beat frequency of the up-chirp stage is $f_{\rm up} = \mu \tau - f_d$. Similarly, it is able to derive the received difference signal of the down-chirp stage as
\begin{align}
  {\bf d}_{r,\rm down}(t) &= {\rm LPF} \left[{\rm Re} \left({\bf y}(t) \right) \cdot {\rm Re} ({\rm e}^{ j \left(2\pi f_c t - \pi \mu t^2\right) }) \right],  \notag \\
  &= {\cal L} {\bf V} {\bf A} {\bf W} {\bf C} \cdot {\rm e}^{j [2 \pi (\mu \tau + f_d) t - 2 \pi (f_c \tau +\frac{1}{2} \mu \tau) \tau) ]} \notag \\&+  {\bf V} {\bf A}_K {\bf I}_K(t) + {\bf n}(t),
\end{align}
with the beat frequency of the down-chirp stage as $f_{\rm down} = \mu \tau + f_d$.

The latency and distance between the BS and TA can be calculated as $\tau = \frac{f_{\rm up} + f_{\rm down} }{\mu}$ and $d_T = \frac {c \tau}{2}$, respectively. The Doppler frequency is equal to $f_d = \frac{f_{\rm down} - f_{\rm up}}{2}$, and the corresponding radial velocity of the TA is $v_T = \frac{\lambda f_d} {2}$.

Obviously, the latency and Doppler frequency can be exactly estimated without estimation error, if the sweep bandwidth and transmit pulse bandwidth satisfy the resolutions of the requirements. However, the matrix ${\bf W}$ and vector ${\bf V}$ affect the received power of TA, how to design ${\bf W}$ and ${\bf V}$ is presented in the following section.

% section-4
\section{A joint beamforming and resource allocation algorithm}\label{section4}
To design the precoding matrix ${\bf W}$ and processing vector ${\bf V}$, this section sets the maximum sum-rate as our objective function. Actually, the sum-rate consists of both communication rate and radar rate. It is important to denote a unified sum-rate of a DFRC system. As known, communication sum-rate (or channel capacity) is widely used in communication systems, which indicates the quality of a communication system. To unify the communication and radar systems, we present a concept of radar rate, similarly to the conception of communication rate. Thus, this paper exploits the maximum sum-rate that includes both communication and radar rates as our objective function.

\subsection{Communication rate and radar rate}
This paper only considers the downlink multiuser channel capacities between the BS and the UEs as the communication rate. Recalling (\ref{comm_signal}), the SINR of the $k$th UE can be expressed as
\begin{equation}\label{SINR}
   \gamma_{k} = \frac{  { P_k \left| {\bf h}_k {\bf w}_k \right|^2  } }
                     {  { \left| \sqrt{P_T} \, {\bf h}_k {\bf w}_T  + \sum_{i=1,i\neq k}^{K} { \sqrt{P_i} \, {\bf h}_k {\bf w}_i } \right|^2 } + \sigma_{k}^2},
\end{equation}
where $ P_k \left| {\bf h}_k {\bf w}_k \right|^2 $ and $\left| \sqrt{P_T} \, {\bf h}_k {\bf w}_T + \sum_{i=1,i\neq k}^{K} { \sqrt{P_i} \, {\bf h}_k {\bf w}_i } \right|^2$ are respectively the power of received signal and interference of the $k$th UE. Thus, the communication rate of all the UEs is equal to
\begin{equation}\label{communication_rate}
  R_{\rm com} = \sum_{k=1}^{K}{\rm log}_2 \left(1 + \gamma_{k} \right),
\end{equation}
where the subscript ``com" of $R_{\rm com}$ stands for communication system.

Whether a TA exists or not, it is determined by the received power of TA at the BS. According to the definition of radar equation \cite{radar_equation}, if the received power of TA is greater than the minimum detection power, i.e. $P_{th}$, the TA can be detected. According to (\ref{SINR_T}), the SINR of the TA is calculated as
\begin{equation}\label{gamma_T}
   \gamma_{T} = \frac{ {\cal E}_{\rm rad} {\bf V} {\bf Z} {\bf W} {\bf W}^{\rm H} {\bf Z}^{\rm H} {\bf V}^{\rm H} } { { {\bf V} {\bf A}_K {\bf A}_K^{\rm H} {\bf V}^{\rm H} } + {\bf V}{\bf V}^{\rm H} \, \sigma^2},
\end{equation}
where ${\bf Z}  = {\cal L} {\bf A} \, {\rm e}^{j (2 \pi f_c \tau + \pi \mu \tau^2)}$, ${\cal E}_{\rm rad} = {\mathbb E} [{\bf C} {\bf C}^{\rm H}] $ is the power of radar transmit pilot sequence, and the term ${\bf V} {\bf A}_K {\bf A}_K^{\rm H} {\bf V}^{\rm H}$ indicates the interferences of other UEs.

Similarly to the concept of communication rate, we define the radar rate by 
\begin{equation}\label{radar_rate}
  R_{\rm rad} = \log_{2} \left( 1 + \gamma_T \right),
\end{equation}
where the subscript ``rad" stands for radar, and the radar rate is affected by ${\bf V}$ and {\bf W}. Obviously, the defined radar rate can reflect the target detection performance, a large $R_{\rm rad}$ stands for a better detection performance.

\subsection{Objective function}
In order to ensure both the communication and radar performances, the optimal function is to maximize the sum-rate of the system, with the constraints given by
\begin{align}\label{objective1}
\mathop {\max} \quad R_{\rm sum} &= \sum_{k=1}^{K}{\rm log}_2 \left(1 + \gamma_{k} \right) + \log_{2} \left( 1 + \gamma_{T} \right) \notag\\
&= {\rm log}_2 {\rm \ Det} \left( {\bf I}_{K+1} + \boldsymbol{\Gamma}_{K+1} \right). \\
\mathop \text{s.t.} \ C1:\ & \gamma_k \geq {2^{\rho_k} - 1}, \forall k, \notag \\
\mathop \text{} \ C2:\ & \gamma_T \geq {2^{\rho_T} - 1}, \notag \\ 
\mathop \text{} \ C3:\ & P_T + \sum_{k=1}^{K} P_k \leq P_{tot},  \notag
\end{align}
where $P_{tot}$ is the total transmit power of the BS, $\rho_k$ and $\rho_T$ are respectively the thresholds of the required SINR of UE and TA. $\boldsymbol{\Gamma}_{K+1} = {\rm Diag}{[\gamma_1,\gamma_2,...,\gamma_K,\gamma_T]}$ is a diagonal matrix.

There are three constraints. $C1$ is to certify the communication the quality of service (QoS) requirement of each UE. $C2$ is used to ensure the TA detection, and the radar rate $R_{\rm rad}$ should be larger than the threshold $\rho_T$. $C3$ is the total power constraint.

Obviously, the optimization issue is non-convex. Therefore, we propose a sub-optimal joint resource allocation algorithm, including two parts: user selection, and precoding \& processing design with power allocation. It is noted that the UE's weight ${\bf w}_k$ is constant, while the TA's weight ${\bf w}_T$ is various during different $M$ blocks. The user selection part is to select $K$ UEs from the user set, so that to maximize $R_{\rm com}$. Based on the selected UEs, the power allocation algorithm is designed to acquire $\bf W$ and $\bf V$ to obtain the sub-optimal solution of (\ref{objective1}).

\subsection{Sub-issue: user selection}
This part is discussed based on a given $\bf W$. We adopt MRT precoding as the initial iteration value, where ${\bf W}_{MRT} = {\bf H}^{\rm H}$ and the subscript ``MRT" indicates maximum ratio transmission (MRT) \cite{MRT_precoding}. Note that the user selection result will not affect the design of $\bf W$. Actually, $\gamma_k$ is mainly determined by each UE's $\theta_k$ and $\varphi_k$. Define $ \Psi = \{q_1, q_2, ..., q_u,... q_U \}$ and $\psi = \{q_{i_1}, q_{i_2}, ..., q_{i_k}, ..., q_{i_K}\} $ by the sets of all the UEs and the selected served UEs respectively, where $1 \le i_1 < i_2 <... <i_K \le U$. Obviously, $\psi$ is a subset of $\Psi$, i.e., $\psi \subseteq \Psi$. The optimal user selection algorithm is traversal algorithm, and its complexity is defined by the size of user set, which is $\mathcal{O} (U^{K})$ complex multiplications. If $U$ and $K$ are small, the traversal algorithm is appreciated. However, when $U$ and $K$ are large, the complexity will increase dramatically. Thus, it is important to present an algorithm to reduce the algorithm complexity.

This paper proposes a low complexity sub-optimal user selection algorithm, named as successive maximum iteration (SMI) user selection algorithm. The main idea behind SMI is to select the UE who can provide the maximum $R_{\rm com}$ during each iteration. Then, the selected UE is removed from the user set, and goes to the next selection step. The iteration is repeated, until all the $K$ UEs are selected. Thus, the complexity of SMI is $\mathcal{O} (UK)$, which is much lower than that of the traversal method. The SMI is shown in Algorithm \ref{SMI}.
\begin{algorithm}[t]
    \caption{Successive maximum iteration user selection algorithm}
    \label{SMI}
    \begin{algorithmic}[1] 
        \Require $\Psi$, $G_u$, $d_{u}$, $\alpha$, $\delta_{u}$, ${\bf a}_t(\theta_{u},\varphi_{u})$ 
        \Ensure $\psi$
        
        \State  \textbf{Initialization} $K$ and $M$.
        \State  \textbf{for} $m = 1: M$
        \State  \quad  ${\bf w}_T =  \mathrm{vec} \left( { {\bf{a}}_{tx}^{\rm{T}}(\theta_m,\varphi_m)} {\bf{a}}_{ty}(\theta_m,\varphi_m) \right) \in \mathbb{C}^{1 \times N_t} $
        \State  \quad \textbf{when} $K < U$    
        \State  \quad \quad Calculate ${\bf h}_{u} = \sqrt{ \left(d_u \big/ d_0 \right)^{-\alpha} \delta_{u}} \, {\bf a}_t (\theta_{u},\varphi_{u})$.
        \State  \quad \quad Let $ {\bf w}_{u} = {\bf h}_{u}^{\rm H} $.
        \State  \quad \quad \textbf{for} $k = 1: K$
	    \State  \quad \quad \quad \textbf{for} $ q_u \in \Psi $
	    \State  \quad \quad \quad \quad Let $\psi = \psi + \{ q_u \}$.
	    \State  \quad \quad \quad \quad Calculate $ c_k = \max \mathop{\sum}\limits_\psi {\rm log}_2 \left( 1 + \gamma_u \right)$.
	    \State  \quad \quad \quad \textbf{end}
	    \State  \quad \quad \quad \textbf{if} $ c_k > c_\Delta $
	    \State  \quad \quad \quad \quad Select the UE $ q_{i_k} = \mathop{ {\rm arg} } c_k$.
	    \State  \quad \quad \quad \quad Update $c_\Delta = c_k$, $\Psi = \Psi - \{ q_{i_k} \}$ and $\psi = \psi + \{ q_{i_k} \}$.
	    \State  \quad \quad \quad  \textbf{else}
	    \State  \quad \quad \quad \quad Output the selected UE set $\psi$.
	    \State  \quad \quad  \textbf{end}
	    \State  \quad  \textbf{end}
	    \State  \textbf{end}
    \end{algorithmic}
\end{algorithm}

\subsection{Sub-issue: precoder and processor design with power allocation}
In the following discussion, assume the UEs have been selected, and the CSI of the selected $K$ UEs are known. Moreover, suppose there exists a TA at the location of the $m$th block time. It is well known that the Lagrange multiplier method (LM) \cite{LM} is one of the most popular algorithm to solve the optimization issue. However, the proposed issue is non-convex, thus, we explore the majorization minimization algorithm (MM) \cite{MM} with the Lagrange multiplier method, i.e. maximum minimization Lagrange multiplier algorithm (MMLM), to solve our problem.

The MM procedure consists of two stages. The first stage is to find a surrogate function, which is simple and solvable. Then, we minimize the gap between the original function and surrogate function at a specific point. Besides, the surrogate function is an upper bound of the objective function. The second stage is to minimize the surrogate function, and find the optimal value, bringing the derived solutions for the next MM iteration. The procedure is repeated until reaching the maximum iteration number.

The key issue of applying MM is to find a surrogate function. In \cite{LM}, it shows that the function ${\rm log \, [Det}({\boldsymbol \Sigma}) ] $ can be given as
\begin{equation}
  {\rm log} \left[{\rm Det} ({\boldsymbol \Sigma}) \right] \leq {\rm log \, [ Det} ({\boldsymbol \Sigma}_0) ] + {\rm tr} \left( \Sigma_0^{-1} ({\boldsymbol \Sigma} - {\boldsymbol \Sigma}_0) \right),
\end{equation}
where ${\boldsymbol \Sigma}_0$ is a constant matrix, and the inequality strictly holds with ${\boldsymbol \Sigma} = {\boldsymbol \Sigma}_0$. The problem of (\ref{objective1}) can reformulate as
\begin{align}\label{objective2}
\mathop {\min} \quad -R_{\rm sum} &= {\rm log}_2 \left[ {\rm Det} \left( {\bf I}_{K+1} + \boldsymbol{\Gamma}_{K+1} \right)^{-1} \right] . \\
\mathop \text{s.t.} \ & C1,\ C2, \ C3. \notag
\end{align}
Thus, (\ref{objective2}) becomes 
\begin{align}\label{upperbound}
  {\rm log}_2 & \, \left[ {\rm Det} \left( {\bf I}_{K+1} + \boldsymbol{\Gamma}_{K+1} \right)^{-1} \right] \leq {\rm log}_2 \, \left[ {\rm Det} \left( {\bf I}_{K+1} + \boldsymbol{\Gamma}_{MRT} \right)^{-1} \right] \notag \\ & + {\rm tr} \left[ ({\bf I}_{K+1}  + \boldsymbol{\Gamma}_{MRT}) (\boldsymbol{\Gamma}_{K+1} - \boldsymbol{\Gamma}_{MRT})^{-1} \right],
\end{align}
where $\boldsymbol{\Gamma}_{MRT}$ is the diagonal matrix as aforementioned.

Our goal is to minimize the right side of (\ref{upperbound}). Since $\boldsymbol{\Gamma}_{MRT}$ is a constant matrix, the problem is equivalent to maximize the trace of $\boldsymbol{\Gamma}_{K+1}$,  as
\begin{align}\label{objective4}
\mathop {\max} \quad {\rm tr}(\boldsymbol{\Gamma}_{K+1}) &= \sum_{k=1}^{K} \gamma_k + \gamma_T. \\
\mathop \text{s.t.} \ C1,\ C2,\ & C3. \notag
\end{align}
Obviously, (\ref{objective4}) is a convex function, which can be solved by Lagrange multiplier method. Define Lagrange function as
\begin{align}\label{lagrange}
  &  L\left({\bf W}, {\bf V}, {\boldsymbol \eta} \right) = \sum_{k=1}^{K} \big[ \left(1 + \eta_k \right) \gamma_k + \eta_k (1 - 2^{\rho_k}) \big] \notag \\ & + (1 + \eta_T) \gamma_T + \eta_T \left(1 - 2^{\rho_T} \right) + \eta_P \bigg( \sum_{k=1}^{K} P_k + P_T - P_{tot} \bigg),
\end{align}
where ${\boldsymbol \eta} = \left[\eta_1,...,\eta_K,\eta_T, \eta_P \right] \in \mathbb{C}^{1 \times (K+2)} $ is a vector of non-positive Lagrange multipliers. The Karush-Kuhn-Tucker (KKT) conditions \cite{Karush-Kuhn-Tucker} of (\ref{lagrange}) are
\begin{align} 
   \ & \frac{\partial L\left({\bf W}, {\bf V}, {\boldsymbol \eta} \right)}{\partial {\bf W}} = {\bf 0}, \tag{22a} \label{KKT1} \\
   \ & \frac{\partial L\left({\bf W}, {\bf V}, {\boldsymbol \eta} \right)}{\partial {\bf V}} = {\bf 0}, \tag{22b} \label{KKT2} \\ 
   \ & {\eta_{\upsilon}} \le 0, {\upsilon} \in \{1,2,...,K,T,P\}, \tag{22c} \label{KKT3}\\ 
   \ & {\gamma_{\upsilon}} + 1 - 2^{\rho_{\upsilon}} \ge 0, \tag{22d} \label{KKT4} \\ 
   \ & {\eta_{\upsilon}} (\gamma_{\upsilon} + 1 - 2^{\rho_{\upsilon}}) = 0, \tag{22e} \label{KKT5} \\
   \ & {\eta_P} \le 0, \tag{22f} \label{KKT6} \\ 
   \ & \sum_{k=1}^{K} P_k + P_T - P_{tot} \le 0, \tag{22g} \label{KKT7}\\ 
   \ & {\eta_P} \bigg( \sum_{k=1}^{K} P_k + P_T - P_{tot} \bigg) = 0, \tag{22h} \label{KKT8}
\end{align}
where (\ref{KKT1}) and (\ref{KKT2}) can be extended as 
\begin{small}
\begin{align}\label{condi_1}
  \frac{\partial L\left({\bf W}, {\bf V}, {\boldsymbol \eta} \right)}{\partial {\bf w}_k} &= 
    \frac{2 (1+{\eta_k}) P_k \, |{\bf h}_k {\bf w}_k |{\bf h}_k}{ { \left| \sqrt{P_T} \, {\bf h}_k {\bf w}_T  + \sum_{i=1,i\neq k}^{K} { \sqrt{P_i} {\bf h}_k {\bf w}_i } \right|^2 } + \sigma_{k}^2 } \notag \\ & - \sum_{i=1 \atop i\neq k}^{K} \frac{2  (1+{\eta_i}) P_i \, |{\bf h}_i {\bf w}_i |^2 |{\bf h}_i {\bf w}_k |{\bf h}_i} { \left({ \left| \sqrt{P_T} \, {\bf h}_i {\bf w}_T + \sum_{j=1, j \neq i}^{K} { \sqrt{P_j} \, {\bf h}_i {\bf w}_j } \right|^2 } \! + \! \sigma_{i}^2 \right)^2 } \notag\\ & + (1+{\eta_T}) \cdot \frac{ {\cal E}_{\rm rad} {\bf V} {\bf Z} \frac{ \partial ({\bf W} {\bf W}^{\rm H}) }{ \partial {\bf w}_k } {\bf Z}^{\rm H} {\bf V}^{\rm H} } { { {\bf V} {\bf A}_K {\bf A}_K^{\rm H} {\bf V}^{\rm H} } + {\bf V}{\bf V}^{\rm H} \sigma^2}, 
\end{align}
\begin{align}\label{condi_2}
    \frac{\partial L\left({\bf W}, {\bf V}, {\boldsymbol \eta} \right)}{\partial {\bf V}} &= \bigg[ \frac{ (2 {\bf Z} {\bf W} {\bf W}^{\rm H} {\bf Z}^{\rm H} {\bf V}^{\rm H}) ({\bf V} {\bf A}_K {\bf A}_K^{\rm H} {\bf V}^{\rm H} + {\bf V}{\bf V}^{\rm H} \, \sigma^2) }  { ({ {\bf V} {\bf A}_K {\bf A}_K^{\rm H} {\bf V}^{\rm H} } + {\bf V}{\bf V}^{\rm H} \, \sigma^2)^2 } \notag \\ &- \frac{({\bf V} {\bf Z} {\bf W} {\bf W}^{\rm H} {\bf Z}^{\rm H} {\bf V}^{\rm H}) (2 {\bf A}_K {\bf A}_K^{\rm H} {\bf V}^{\rm H} + 2 {\bf V}^{\rm H} \, \sigma^2)}{ ({ {\bf V} {\bf A}_K {\bf A}_K^{\rm H} {\bf V}^{\rm H} } + {\bf V}{\bf V}^{\rm H} \, \sigma^2)^2 } \bigg] \notag \\ & \cdot (1+{\eta_T}) \, {\cal E}_{\rm rad}.
\end{align}
\end{small}

To obtain the solutions of ${\bf W}$ and ${\bf V}$ in (22), we adopt the alternating direction method of multipliers (ADMM). The initialized ${\bf W}$ by MRT is denoted by ${\bf W}^{(0)}$, where ``0" indicates the initial step. Let $\frac{\partial L\left({\bf W^{(0)}}, {\bf V}, {\boldsymbol \eta} \right)}{\partial {\bf V}} = {\bf 0}$, then it is derived that
\begin{align}\label{v_value}
  {\bf Z} {\bf W}^{(0)} ({\bf W}^{(0)})^{\rm H} {\bf Z}^{\rm H} {\bf V}^{\rm H} \left({\bf V} {\bf A}_K {\bf A}_K^{\rm H} {\bf V}^{\rm H} + {\bf V}{\bf V}^{\rm H} \, \sigma^2 \right) \notag & =  \\ {\bf V} {\bf Z} {\bf W}^{(0)} ({\bf W}^{(0)})^{\rm H} {\bf Z}^{\rm H} {\bf V}^{\rm H} \left({\bf A}_K {\bf A}_K^{\rm H} {\bf V}^{\rm H} + {\bf V}^{\rm H} \, \sigma^2 \right).
\end{align}
By solving (\ref{v_value}), we can obtain ${\bf V}^{(1)}$ and then bring it to (\ref{KKT1}).

Set $\frac{\partial L\left({\bf W}, {\bf V}^{(1)}, {\boldsymbol \eta} \right)}{\partial {\bf w}_k } = {\bf 0}$, and achieve ${\bf W}^{(1)}$. According to the derived ${\bf V}^{(1)}$ and ${\bf W}^{(1)}$, do water filling power allocation algorithm to maximum the trace
\begin{small}
\begin{align}
  {\rm tr} \Big(\boldsymbol{\Gamma}_{K+1}^{(1)} \Big)  \!&=  \!\sum_{k=1}^{K}{ \frac{ { P_k \left| {\bf h}_k {\bf w}_k^{(1)} \right|^2  } } {  { \left| \sqrt{P_T} \, {\bf h}_k {\bf w}_T^{(1)}  \!+ \! \sum_{i=1,i\neq k}^{K} { \sqrt{P_i} \, {\bf h}_k {\bf w}_i^{(1)} } \right|^2 } \! + \!\sigma_{k}^2 }} \notag
  \\ &+ \frac{ {\cal E}_{\rm rad} {\bf V}^{(1)} {\bf Z} {\bf W}^{(1)} {{\bf W}^{(1)}}^{\rm H} {\bf Z}^{\rm H} { {\bf V}^{(1)}}^{\rm H} } { { {\bf V}^{(1)} {\bf A}_K {\bf A}_K^{\rm H} {{\bf V}^{(1)}}^{\rm H} } + {\bf V}^{(1)} {{\bf V}^{(1)}}^{\rm H} \sigma^2},
\end{align}
\end{small}and the optimal power allocation results $P_k^{(1)}$ and $P_T^{(1)}$ can be obtained.

Following, take ${\bf V}^{(1)}$ and ${\bf W}^{(1)}$ into (\ref{upperbound}), and obtain the next surrogate function. By optimizing the new surrogate function, we can solve ${\bf V}^{(2)}$ and ${\bf W}^{(2)}$ by KKT conditions, and then do water filling power allocation to maximize ${\rm tr} \big(\boldsymbol{\Gamma}_{K+1}^{(2)} \big)$. Repeat this process, until the iteration number $i$ reaches the maximum iteration number $\nu_{max}$, or satisfying $ \Vert {\bf W}^{(i)} - {\bf W}^{(i-1)} \Vert^2 < \varepsilon $ and $ \Vert {\bf V}^{(i)} - {\bf V}^{(i-1)} \Vert^2 < \varepsilon $, where $\varepsilon$ is a small position number, i.e., $10^{-3}$. Finally, we can obtain the solutions ${\bf W} = {\bf W}^{(\nu_{max})}$ and ${\bf V} = {\bf V}^{(\nu_{max})}$, and the power allocation results $P_k^{(\nu_{max})}$ and $P_T^{(\nu_{max})}$, where $\nu_{max}$ is the final iteration times. The MMLM is presented in Algorithm \ref{MMLM}.

\begin{algorithm}[t]
    \caption{Maximum minimization Lagrange multiplier algorithm for designing {\bf W} and {\bf V}}
    \label{MMLM}
    \begin{algorithmic}[1] 
        \Require $\psi$, $\gamma_k$, $\gamma_T$, $\rho_k$, $\rho_T$, $\varepsilon$, $\nu_{max}$
        \Ensure ${\bf W}$ and ${\bf V}$
        
        \State \textbf{Initialization} ${\bf W}^{(0)}$ by MRT precoding.
        \State Set $i=1$, $\varepsilon$ and $\nu_{max}$.
        \State \textbf{while} $i < N_{ite}$ \& $ \Vert {\bf W}^{(i)} - {\bf W}^{(i-1)} \Vert^2 \ge \varepsilon $ \&  $ \Vert {\bf V}^{(i)} - {\bf V}^{(i-1)} \Vert^2 \ge \varepsilon $
        \State \quad Find the surrogate function according to (\ref{upperbound}) and optimize the surrogate function.
        \State \quad Construct Lagrange function $L\left({\bf W}^{(i)}, {\bf V}^{(i)}, {\boldsymbol \eta} \right)$.
        \State \quad Calculate KKT conditions of (\ref{lagrange}).
        \State \quad Calculate ${\bf V}^{(i)}$ by $\frac{\partial L\left({\bf W}^{(i-1)}, {\bf V}^{(i)}, {\boldsymbol \eta} \right)}{\partial {\bf V}^{(i)}} = {\bf 0}$ and bring it back to (\ref{condi_1}).
        \State \quad Calculate ${\bf w}_k^{(i)}$ by $\frac{\partial L\left({\bf W}^{(i)}, {\bf V}^{(i)}, {\boldsymbol \eta} \right)}{\partial {\bf w}_k^{(i)}} = {\bf 0}$ and update KKT conditions. 
        \State \quad Do water filling algorithm to maximize the trace ${\rm tr} \big(\boldsymbol{\Gamma}_{K+1}^{(i)} \big)$, acquire the optimal power allocation results $P_k^{(\nu_{max})}$ and $P_T^{(\nu_{max})}$.
        \State \textbf{end} 
        \State ${\bf W} = {\bf W}^{(\nu_{max})}$ and ${\bf V} = {\bf V}^{(\nu_{max})}$
    \end{algorithmic}
\end{algorithm}

As a summary, the joint resource allocation algorithm includes user selection, and precoder and processor design with power allocation. During the initialization, we firstly do SMI to find the UE set $\psi$, then utilize MMLM to calculate ${\bf W}$, ${\bf V}$, $P_T$ and $P_k$. Overall, the whole algorithm needs at least $(UK \nu_{max})$ complex multiplications.

% section-5
\section{Theoretical analysis}\label{section5}
In this section, we derive the spectrum, BER and AF of the proposed system.

\subsection{Spectrum}
Actually, the spectrum of the integrated signal is determined by the transmit signal $s_{n_t}(t)$. Since the spectrum of the bandpass and baseband signals are equivalent, we only consider the equivalent baseband signal $\hat s_{n_t}(t)$ of ({\ref{CPM-BF-LFM}}), which is given by
\begin{align}\label{baseband}
   {\widehat{s}}_{n_t}(t) = &\sum_{n=1}^N x_{n_t,n} \, {g \left( t-(n-1)T_s \right)} \, {\rm e}^{ j \pi \mu t^2}, \notag \\ 
    = &\sum_{n=1}^N \Big( w_{n_t,T} \, {\rm e}^{j \sum_{i=1}^{n} b_{T,i} h \pi} + \sum_{k=1}^K w_{n_t,k} \, {\rm e}^{j \sum_{i=1}^{n} b_{k,i} h \pi} \Big) \,\notag \\ & \cdot {g \left( t-(n-1)T_s \right)} \, {\rm e}^{ j \pi \mu t^2}.
\end{align}

Since the spectrum of different transmit antennas are the same, we take the $n_t$th antenna as an example for analyzing. Define the Fourier transform of $\widehat{s}_{n_t}(t)$ by $ {\mathcal S}(f) $, which can be expressed as

\begin{small}
\begin{align}\label{spectrum}
    \mathcal{S}(f) &=  \!\mathlarger{\int}_{-\infty}^{+\infty} \sum_{n=1}^N \Big( w_{n_t,T} \, {\rm e}^{j \sum_{i=1}^{n} b_{T,i} h \pi}  \!+ \! \sum_{k=1}^K w_{n_t,k} \, {\rm e}^{j \sum_{i=1}^{n} b_{k,i} h \pi} \Big) \, \notag \\ & \cdot {g \left( t-(n-1)T_s \right)} \, {\rm e}^{ j \pi (\mu t^2 - 2 f t)} \,{\rm d} t, \notag
    \\  &= \sum_{n=1}^N \Big( w_{n_t,T} \cdot \prod_{i=1}^{n}  {\rm e}^{j b_{T,i} h \pi} + \sum_{k=1}^K w_{n_t,k} \cdot \prod_{i=1}^{n} {\rm e}^{j b_{k,i} h \pi} \Big) \notag \\ & \cdot \int_{0}^{T_B} {g (t)} \, {\rm e}^{ j \pi (\mu t^2 - 2 f t)} \,{\rm d} t.
\end{align}
\end{small}

\begin{comment}
From (\ref{spectrum}), we know that the spectrum is determined by amplitude and bandwidth.
\end{comment}
As mentioned in Section III, $b_{k,i}$ and $b_{T,i}$ are complex signals. Actually, the $n$th symbol of the $n_t$th transmit antenna $x_{n_t,n}$ is a random variable, thus the mean value of $x_{n_t,n}$ is

\begin{small}
\begin{align}\label{mean}
   \mathbb{E} \left[ \left| {x_{n_t,n}} \right| \right] &= \mathbb{E} \left[ \vert w_{n_t,T} \, \prod_{i=1}^{n} {\rm e}^{j b_{T,i} h \pi} + \sum_{k=1}^K w_{n_t,k} \, \prod_{i=1}^{n} {\rm e}^{j b_{k,i} h \pi} \vert \right], \notag \\
   &= \left\vert w_{n_t,T} + \sum_{k=1}^K w_{n_t,k} \right\vert \, \frac{n}{\cal M} \sum_{ \substack{ { \xi= -({\cal M}-1) } \\ {\xi \; is \; odd}} }^{({\cal M}-1) } {\rm e}^{j \frac{\xi}{\cal M} h \pi},
\end{align}
\end{small}where $\xi$ represents one of the ${\cal M}$ values of the modulated independent symbols $b_{k,i}$ and $b_{T,i}$, and the prior probability of which is $\frac{1}{{\cal M}} $. From (\ref{mean}), it is found that different precoding schemes ${\bf W}$ and modulation orders ${\cal M}$ can affect the amplitude of the spectrum. Compared with MRT precoding, the spectrum of our scheme has a larger amplitude, and a smaller ${\cal M}$ derives a lower amplitude. Assume
\begin{align}\label{bandwidth}
  {\mathcal G}(f) &= \int_{0}^{T_B} {g (t)} \, {\rm e}^{ j \pi (\mu t^2 - 2 f t) } \,{\rm d} t, \notag \\ &= {\mathcal G}_u(f) * {\mathcal G}_{LFM}(f),
\end{align}
where ${\mathcal G}_u(f)$ is the Fourier transform of transmit pulse $g(t)$, and $B_g$ is the bandwidth of $g(t)$. If we adopt rectangular pulse, i.e., $B_g = 1/T_s$ or raised-cosine pulse i.e., $B_g = (1 + \alpha_g) /2 T_s$, where $\alpha_g$ is the roll-off factor.

Define ${\mathcal G}_{LFM}(f)$ by the Fourier transform of LFM signal, given by
\begin{align}\label{Fourier}
    {\mathcal G}_{LFM}(f) &=  \int_{0}^{T_B} {\rm e}^{ j \pi (\mu t^2 - 2 f t)} \,{\rm d} t, \notag \\
        &=  {\rm e}^{-\frac{j \pi f^2} {\mu} } \int_{ -\frac{f} {\sqrt{\mu}} }^{ \sqrt{\mu} \,T_B -\frac{f} {\sqrt{\mu}} } \left[ \cos(\pi {t^{'}}^2 ) + j \sin(\pi {t^{'}}^2 ) \right] {\rm d} t^{'},
\end{align}
where $t^{'} = \sqrt{\mu} t - \frac{f} {\sqrt{\mu}} $, $C(x) = \int_{0}^{x} \cos(\pi \alpha^2)  \,{\rm d} \alpha $ and $S(x) = \int_{0}^{x} \sin(\pi \alpha^2)  \,{\rm d} \alpha$ are cosine and sine Fresnel integral respectively. Since both $C(x)$ and $S(x)$ are odd function, the amplitude of $\mathcal{G}_{LFM}(f)$ can be expressed in (\ref{G-LFM-3}).
\begin{comment}
\begin{figure*}[b]
\setcounter{equation}{45}
\hrulefill
\begin{equation}\label{G-LFM-2}
{\mathcal G}_{LFM}(f) = {\rm e}^{- \frac{j \pi f^2}{\mu} } \left[ C\left(\sqrt{\mu} \, T_B -\frac{f} {\sqrt{\mu}} \right) + C\left(\frac{f} {\sqrt{\mu}} \right) + j S\left(\sqrt{\mu} \, T_B -\frac{f} {\sqrt{\mu}} \right) + j S\left(\frac{f} {\sqrt{\mu}} \right) \right].
\end{equation}
\end{figure*}
\end{comment}
\begin{figure*}[t]
\setcounter{equation}{32}
\begin{equation}\label{G-LFM-3}
\Big\vert {\mathcal G}_{LFM}(f) \Big\vert = \sqrt{\left[ C\left(\sqrt{\mu} \, T_B -\frac{f} {\sqrt{\mu}} \right) + C\left(\frac{f} {\sqrt{\mu}} \right) \right]^2 + \left[ S\left(\sqrt{\mu} \, T_B -\frac{f} {\sqrt{\mu}} \right) + S\left(\frac{f} {\sqrt{\mu}} \right) \right]^2}.
\end{equation}
\hrulefill
\end{figure*}

According to (\ref{bandwidth}), it is found that the integrated signal bandwidth is equal to $B = B_w + B_g$, i.e. $B = \mu T_B + 1/T_s$ for rectangular pulse or $B = \mu T_B + (1 + \alpha_g) /2 T_s$ for raised-cosine pulse. To improve the spectrum efficiency, we can adopt bandwidth-saving pulse.

\subsection{BER}
Actually, the BER of the proposed system is independent of LFM. The optimum detector of a CPM signal is realized by Viterbi algorithm \cite{Viterbi}. Considering the proposed system, the BER of the $k$th UE becomes
\begin{align}\label{BER1}
    P_{e,k} &= 2 \int_{0}^{+\infty} { Q \left(\sqrt {\log_2 {\mathcal M} \left(1- \frac{\sin 2 \pi h}{2 \pi h} \right) \gamma_k } \right) \, p(\gamma_k) \, {\rm d} \gamma_k, }
\end{align}
where $Q(x) = \int_{x}^{+\infty} \frac{1}{\sqrt{2 \pi}} \, {\rm e}^{-\frac{t^2}{2} } \, {\rm d}t $, $p(\gamma_k)$ is the probability density function (PDF) of $\gamma_k$. For mathematic convenience, we utilize ${\bf W}^{(0)}$ that is obtained by MRT instead of the iteration optimal ${\bf W}^{(\nu_{max})}$ to calculate $p(\gamma_k)$. The expression of $\gamma_k$ in (\ref{SINR}) is
\begin{equation}
    \gamma_{k} \leq \frac{ { P_k \, \left(d_k \big/ d_0 \right)^{-\alpha} \delta_k \, N_t } }
                 {  { \left| \sqrt{P_T}\, {\bf h}_k {\bf w}_T  + \sum_{i=1,i\neq k}^{K} { \sqrt{P_i}\, {\bf h}_k {\bf w}_i } \right|^2 } + \sigma_k^2 },
\end{equation}
where the path loss $d_k^{-\alpha}$ and shadowing effect $\delta_k$ obey power law and log-normal distributions, the PDF of $\delta_k$ is $p(\delta_k) = \frac{1}{\delta_k \sqrt{2 \pi} \sigma_{\delta_k}  } {\rm e}^{-\frac{{\rm ln}(\delta_k)^2}{2 \sigma_{\delta}} }$, where $\sigma_{\delta}$ is the variance of $\delta_k$.

Let $I_k = {\sqrt{P_T}\, {\bf h}_k {\bf w}_T + \sum_{i=1,i\neq k}^{K} {\sqrt{P_i}\, {\bf h}_k {\bf w}_i } }$ be the interferences, and it is the sum of several independent and identically (i.i.d) distributed random variables. According to the central limit theorem, $I_k$ can be approximately viewed as complex Gaussian distribution, i.e., ${\mathcal N}(\mu_I, \sigma_I)$, which is simulated by Monte Carlo. The mean value of $I_k$ can be derived by $\mu_I = \sqrt{P_T} \, \widetilde{\mu}_T + \sum_{k=1}^{K} \sqrt{P_i} \, \widetilde{\mu}_k $, where $\widetilde{\mu}_T$ and $\widetilde{\mu}_k$ are respectively the mean value of ${\bf h}_k {\bf w}_T$ and $ {\bf h}_k {\bf w}_i $. It can be seen that a larger $K$ leads to a larger mean value $\mu_I$ when the power allocation factor is fixed. Moreover, for a given $K$, e.g., $K=4$, $P_T$ shows a significant influence on the $\mu_I$, indicating that the interference increases with the increased of $P_T$. The variance of $I_k$ can be expressed as $\sigma_I = P_T \, \widetilde{\sigma}_T + \sum_{k=1}^{K} P_i \, \widetilde{\sigma}_k $, where $\widetilde{\sigma}_T$ and $\widetilde{\sigma}_k$ are respectively the variance of interference ${\bf h}_k {\bf w}_T$ and ${\bf h}_k {\bf w}_i$. Thus, the total interference plus noise obeys chi-square distribution with degree of freedom 2, given by
\begin{equation}
    p(I_k) = \frac{1}{2 \sigma_I^2} \, {\rm e}^{-\frac{I_k}{2 \sigma_I^2}}.
\end{equation}
\begin{comment}
\begin{figure}[t]
\centering
\includegraphics[width = 3.0 in]{PDF.pdf}
\caption {The PDF curves versus different $K$ and $P_T$.}
\label{PDF}
\end{figure}
\end{comment}

Assume $z_k = \frac{d_k^{-\alpha} \delta_k}{I_k} $, according to the relationship of product and quotient of random variables, the PDF of $\gamma_k$ is
\begin{equation}\label{gamma_1}
	p(\gamma_k) = \frac{ (2 (\sigma_I+\sigma_k)^2)^{-\frac{1}{\alpha} }  \, {\rm e}^\frac{\sigma_\delta^4}{\alpha^2}  } {\alpha (d_{k,2}-d_{k,1})} \,  \, Q \left( -\frac{\sigma_\delta}{2} \right) \, \Gamma \left(1- \frac{1}{\alpha} \right) \, \gamma_k^{-\frac{1}{\alpha} -1},
\end{equation}
where $\Gamma(x) = \int_{0}^{+\infty} t^{x-1} {\rm e}^{-t} \, {\rm d} t $, $d_{k,1}$ and $d_{k,2}$ are respectively the minimum and maximum range from the BS to the $k$th UE, i.e. $d_{k,1} \le d_k \le d_{k,2}$.

Taking (\ref{gamma_1}) into (\ref{BER1}), we can derive the upper bound of the BER expression. Unfortunately, the closed-form solution can not be deducted since the integral of the $Q$ function can be expressed by the analytical form.

\subsection{AF}
The ambiguity function is defined as the time-frequency response observed at the matched filter of the BS receiver, which is given by \cite{AF}
\begin{equation}\label{AF}
    \left|\chi(\tau, f_d) \right|^2 = \left|\int_{-\infty}^{+\infty} \widetilde{s}_{n_t}(t) \, {\widetilde{s}_{n_t}}^{\,*}(t-\tau) \, {\rm e}^{j 2 \pi f_d t} \, {\rm d} t \right|^2,
\end{equation}
where $\widetilde{s}_{n_t}(t)$ is the complex envelope of baseband signal $\widehat{s}_{n_t}(t)$, substituting ({\ref{baseband}}) into ({\ref{AF}}) yields ({\ref{AF_2}}).
\begin{figure*}[t]
\setcounter{equation}{40}
\begin{equation}
\label{AF_2}
\left|\chi(\tau, f_d) \right|^2 = \left| {\rm e}^{ - j \pi \mu \tau^2} \sum_{n=1}^{N} \, \sum_{i=1}^{N} \, \int_{-\infty}^{+\infty} x_{n_t,n} \, x^{*}_{n_t,i} \,  {g \left( t - (n-1) T_s \right)} \, {g^{*} \left( t - (i-1) T_s -\tau \right)} \, {\rm e}^{ j 2 \pi (\mu \tau + f_d)t } \,{\rm d} t \right|^2.
\end{equation}
\hrulefill
\end{figure*}

Let $t_1 = t - (n-1) T_s$, $q = n-i$, (\ref{AF_2}) becomes
\begin{align}\label{AF_3}
    \left|\chi(\tau, f_d) \right|^2 &=  \bigg| {\rm e}^{ - j \pi \mu \tau^2} \sum_{n=1}^{N} \,  {\rm e}^{ j 2 \pi (\mu \tau + f_d) (n-1) T_s} \notag \\ &\cdot \sum_{i=1}^{N} \,  \chi_1(\tau-q T_s, f_d + \mu \tau) \bigg|^2,
\end{align}
where 
\begin{align}\label{AF_4}
& \chi_1(\tau-q T_s, f_d + \mu \tau) = \int_{-\infty}^{+\infty} x_{n_t,n} \, x^{*}_{n_t,i} \notag \\ & {g \left( t_1 \right)} \, {g^{*} \left( t_1 - (\tau - q T_s) \right)} \, {\rm e}^{ j 2 \pi (\mu \tau + f_d)t_1 } \,{\rm d} t_1,
\end{align}
is the single pulse auto-correlation function.

Let $g(t)$ be a rectangular pulse, regarding as the symmetry property, i.e., $\left|\chi(\tau, f_d) \right|^2 = \left|\chi(-\tau, -f_d) \right|^2$, (\ref{AF_4}) becomes
\begin{align}\label{AF2}
    \left|\chi(\tau, f_d) \right|^2 &= \Bigg| \sum_{q=-(N-1)}^{(N-1)} \frac{\sin \left[ \pi (f_d + \mu \tau) (N-|q|) T_s \right]} {\sin \left[\pi  (f_d + \mu \tau) T_s \right] } \notag \\ & \cdot  \chi_1 \left(\tau-q T_s, f_d + \mu \tau \right) \Bigg|^2 \! \!,
\end{align}

Obviously, the AF is a two-dimensional function of $\tau$ and $f_d$, whose maximum value occurs at $(\tau, f_d) = (0,0)$. The cut along in Doppler axis is by setting $\tau =0$, indicating the resolution in the Doppler dimension, which is
\begin{align}\label{AF3}
    \left|\chi(0, f_d) \right|^2 &= \Bigg| \sum_{q=-(N-1)}^{(N-1)} \! \! \!  \! \frac{\sin \left[ \pi f_d (N-|q|) T_s \right]} {\sin \left(\pi f_d T_s \right) } \chi_1 \left(-q T_s, f_d \right) \Bigg|^2 \! \!.
\end{align}

Set $\left|\chi(0, f_d) \right|^2 = 0 $, the first null in Doppler domain occurs at $ f_{d,w} = \frac{1}{N T_s}$. Similarly, the cut along in latency axis is obtained by setting $f_d =0$, indicating the resolution in the latency dimension, which is
\begin{equation}\label{AF4}
    \left|\chi(\tau, 0) \right|^2 = \left|\sum_{q=-(N-1)}^{(N-1)} \! \! \!  \! \frac{\sin \left[ \pi \mu \tau (N-|q|) T_s \right]} {\sin \left(\pi \mu \tau T_s \right) } \, \chi_1 \left(\tau-q T_s, \mu \tau \right) \right|^2 \! \!.
\end{equation}

Let $\left|\chi(\tau, 0) \right|^2 = 0 $, the first null in latency domain occurs at $ \tau_w = \frac{1}{\mu N T_s} = \frac{1}{B_w}$. It can be seen that $\tau_w$ is completely determined by the LFM sweep bandwidth. Define $\kappa$ by the compression ratio of the transmit and received pulse duration, which is
\begin{equation}\label{product}
    \kappa = \frac{T_B}{\tau_w} = T_B \, B_w.
\end{equation}
$\kappa$ is the multiplication of block time and sweep bandwidth, which is also called time-bandwidth product \cite{time-bandwidth product}. The minimum detection distance is denoted by $d_{\min} = \frac{c}{2 B_w}$, indicating the range resolution, and $c$ is the speed of light. Similarly, the minimum detection velocity is defined by $v_{\min} = \frac{\lambda}{2 T_B}$, reflecting the Doppler resolution.

According to $d_{\min}$ and $v_{\min}$, the time-bandwidth product increases with the increased sweep bandwidth, which is different from the single carrier radar signal. Evidently, a large $\kappa$ can meet both the requirements of the range resolution and Doppler resolution. 

\begin{comment}
\subsection{Detection \& false alarm probabilities}
In the radar system, the target detection probability $P_d$ is defined as the probability that the TA is successfully detected by the radar receiver when it exists; meanwhile, the false alarm probability $P_{fa}$ is the probability that the TA is deceptive or erroneous detected, when there is no TA. According to \cite{radar-signal}, the relationship between the target detection probability and false alarm probability is 
\begin{equation}
  P_d = \frac{1}{2} \, {\rm erfc} \left( \sqrt{-{\rm log} \, P_{fa} } - \sqrt{\gamma_T + 0.5} \right).
\end{equation}  
Thus, the SINR of the TA can be expressed as
\begin{align}
    \gamma_T &= \left[ \sqrt{-{\rm log} \, P_{fa} } - {\rm arg} \, {\rm erfc} \, (2 P_d) \right]^2 - 0.5.
\end{align}
For a given $\gamma_T$, $P_{fa}$ and $P_d$ is a trade-off, a large $P_{fa}$ leads to a better $P_d$, which is benefit for TA detection.
\end{comment}

% section-6
\section{Simulation results}\label{section6}
In this section, we evaluate the performance of our proposed system through numerical simulations. The equipped transmit and received antennas of the BS are respectively $16$ and $4$, which are respectively structured by a $4 \times 4$ and $2 \times 2$ URA. All the UEs are uniform distribution in the cell, and the simulation parameters are listed in Table \ref{parameters}.
\begin{table}[t]
    \caption{Simulation parameters}
    \vspace{1.5pt}
    \centering
    \begin{tabular}{ p{4cm}<{\centering} p{1.5cm}<{\centering} p{1.5cm}<{\centering} }
        \hline
          Significance    &   Parameters    &    Values    \\
        \hline
          Number of UEs                     &   $ K   $         &     $  4   $            \\
          Number of transmit antennas       &   $ N_t   $       &     $  16  $            \\
          Number of received antennas       &   $ N_r   $       &     $  4   $            \\
          Number of blocks                  &   $ M $           &     $ 324  $            \\
          Number of symbols per block       &   $ N $           &     $ 20   $            \\
          Symbol duration                   &   $ T_s $         &     $  5 $ $\mu$s       \\       
          Carrier Central frequency         &   $ f_c $         &     $ 2.4 $ GHz         \\
          Chirp rate                        &   $ \mu  $        &     $ 10^{10} $         \\
          Pass loss factor                  &   $ \alpha $      &     $ 3.0 $             \\ 
          Transmit power                    &   $ P_{tot} $     &     $ 30 $ dBm          \\ 
          Reference range                   &   $ d_0 $         &     $ 100 $ m           \\ 
        \hline
    \end{tabular}
    \label{parameters}
\end{table}

\begin{figure*}[t]
\centering
\includegraphics[width = 7.1 in]{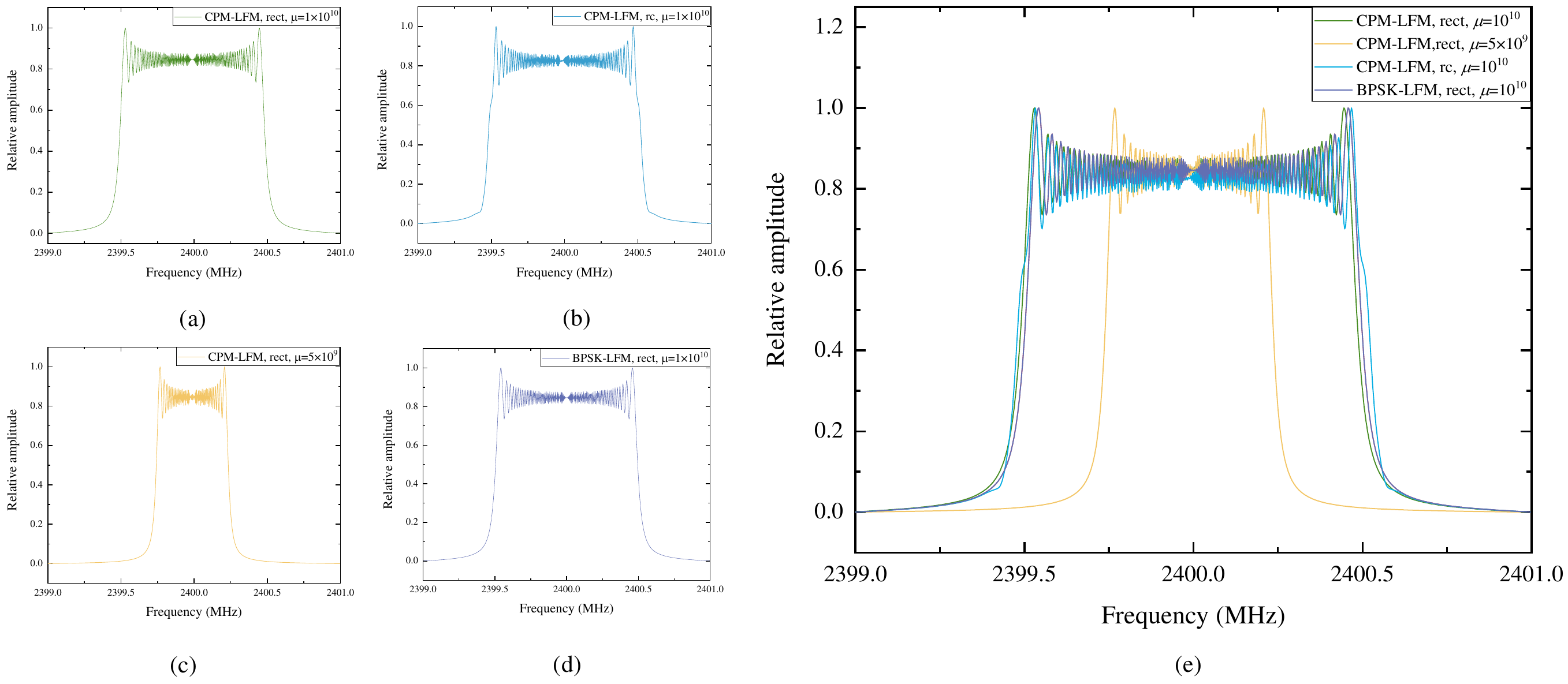}
\caption {Spectrum of the integrated signals with various chirp rates and pulses, where the subscript ``rect" and ``rc" represent the rectangular and raised-cosine pulse. (a) CPM-LFM signal with rectangular pulse, where $\mu = 1 \times 10^{10}$; (b) CPM-LFM signal with raised-cosine pulse, where $\mu = 1 \times 10^{10}$; (c) CPM-LFM signal with rectangular pulse, where $\mu = 5 \times 10^{9}$; (d) BPSK-LFM signal with rectangular pulse, where $\mu = 1 \times 10^{10}$; (e) overall spectrum comparison. }
\label{spectrum_signal}
\end{figure*}

\subsection{Spectrum}
Fig. 4 shows the spectrum of the proposed integrated signal, with various chirp rates and pulses. It can be seen that the bandwidth of integrated waveforms are determined by both sweep bandwidth $B_w$ and transmit pulse bandwidth $B_g$, independent of the center frequency $f_c$. For a given $\mu$, if $B_w$ is small, the ripples are very evident. On contrast, the spectrum tends to be a rectangular profile with the increasing of $B_w$. Obviously, the bandwidth of a raised cosine pulse is smaller than that of a rectangular pulse, for a given $T_s$. Moreover, the chirp rate $\mu$ affects the bandwidth significantly, a larger $\mu$ may lead to a larger bandwidth, e.g., the bandwidth of the curve in $\mu = 5 \times 10^{9}$ is half of the $\mu = 1 \times 10^{10}$. Furthermore, for a given $g(t)$ and $\mu$, the bandwidth of CPM is narrower than that of BPSK. From the spectrum, it is found that ${\bf W}$ has no effect to the bandwidth of the proposed system, verifying the analysis.
% Here is Figure-4 spectrum

\subsection{BER}
\begin{figure}[t]
\centering
\includegraphics[width = 3.2 in]{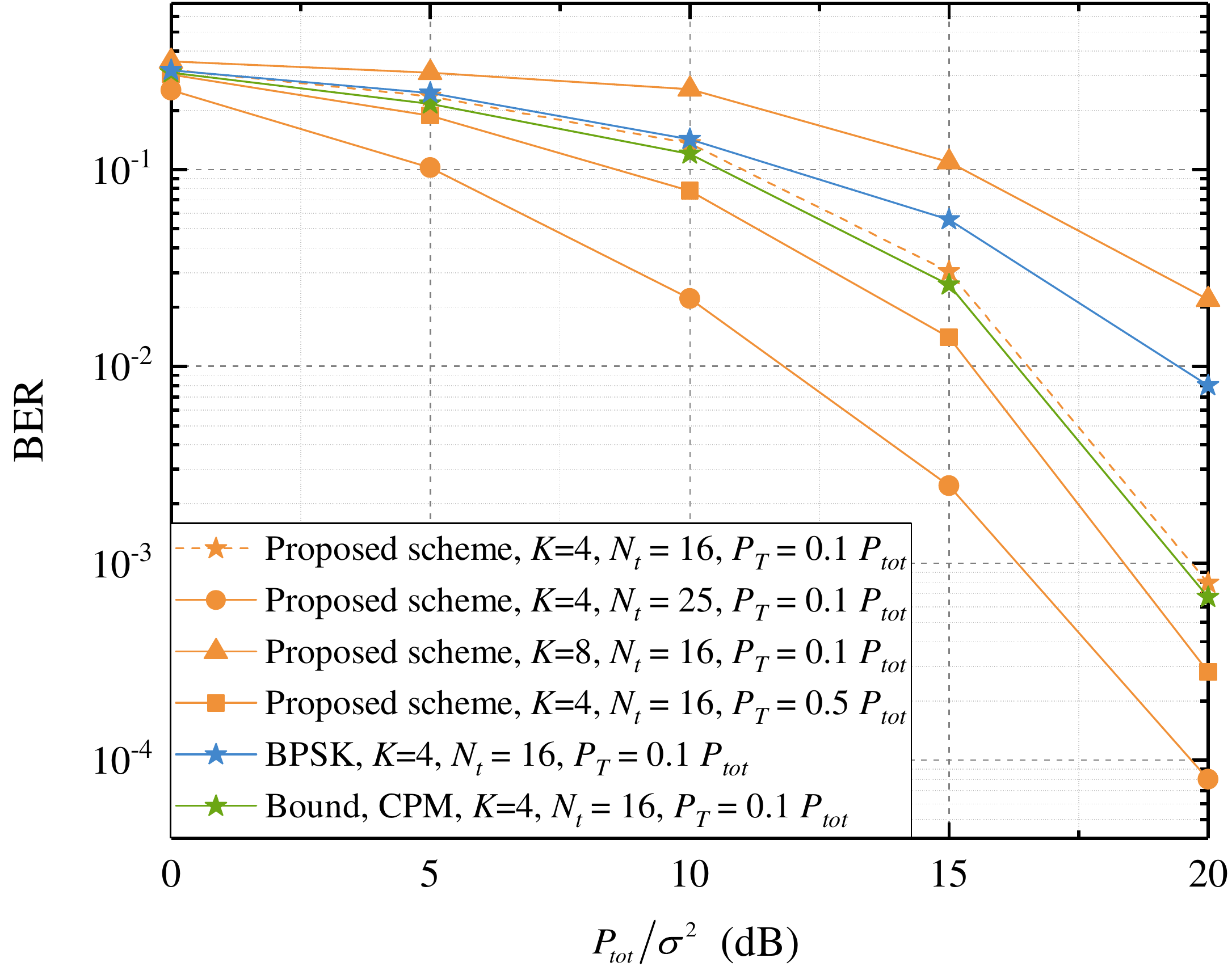}
\caption {The average BER versus $P_{tot}/{\sigma^2}$ in different modulation types, precoding schemes, number of UEs and number of transmit antennas.}
\label{BER_performance}
\end{figure}

The average BER performance is shown in Fig. \ref{BER_performance}. When $K=4$, $N_t=16$ and $P_T = 0.1 P_{tot}$, it can be seen that the proposed scheme achieves $10^{-3}$ when SNR is $20$ dB. The theoretical curve is the upper bound, since $\gamma_k$ can get the maximum value. Moreover, when $N_t$ is given, i.e., $N_t=16$, a large $K$ results in a worse BER performance because of the remaining interferences. In terms of the number of transmit antennas, a large $N_t$ can eliminate the interferences effectively, providing a better BER performance. Furthermore, power allocation result also has an impact on the BER performance, since a larger $P_T$ leads to the reduction of communication power $P_k$, indicating a deteriorated BER. As expected, when $K$, $N_t$ and $P_T$ are fixed, the proposed scheme provides better BER performance than BPSK counterpart, since the Viterbi algorithm is exploited for the CPM signal.
% Here is Figure-5 BER

\subsection{AF}
\begin{figure}[t]
\centering
\includegraphics[width = 3.2 in]{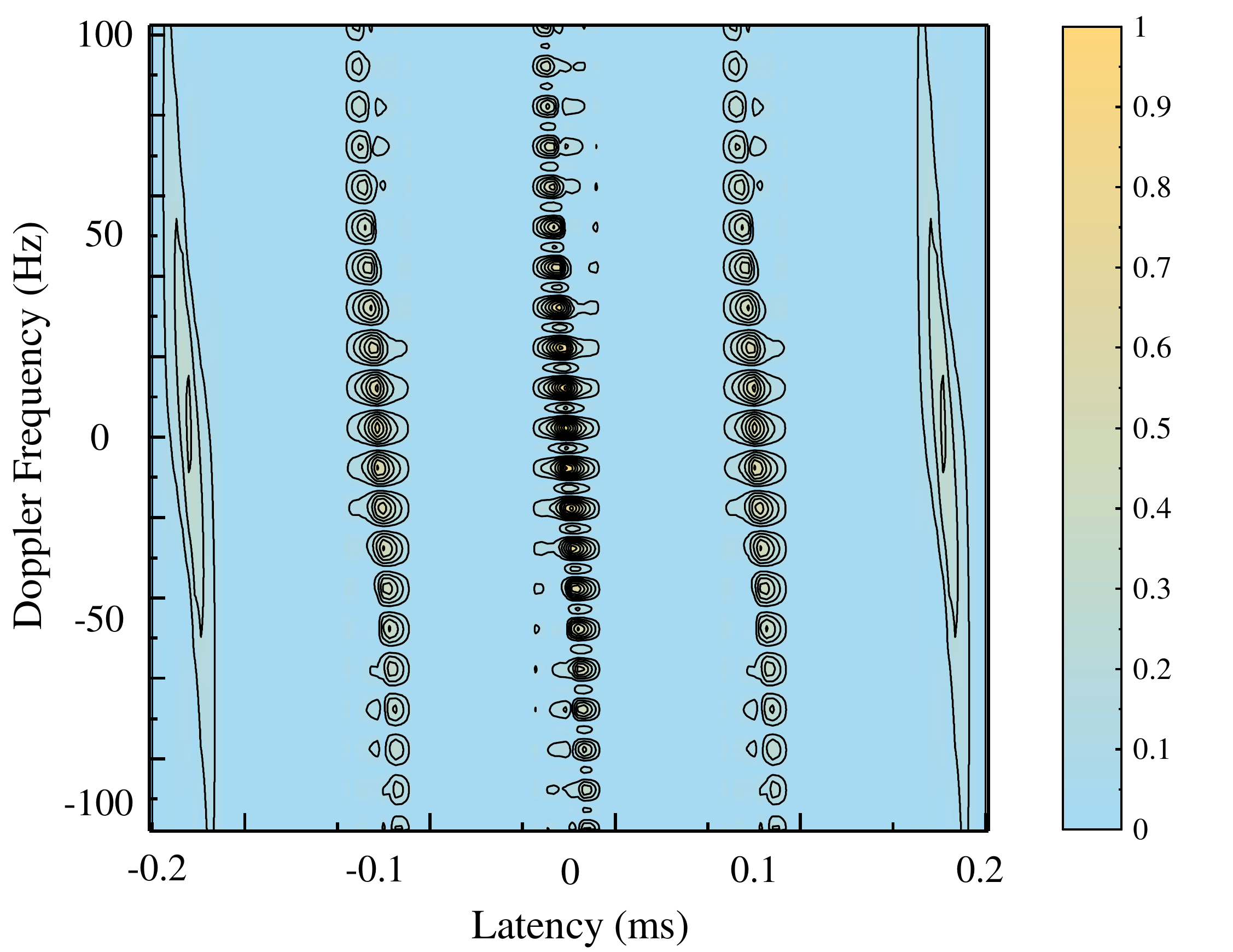}
\caption {The AF contour of our proposed system in latency and Doppler frequency domains, where $g(t)$ is rectangular pulse. }
\label{AF_contour}
\end{figure}

The AF of our proposed system is shown in Fig \ref{AF_contour}, the horizontal and vertical axes stand for latency and Doppler frequency domains. When there is no TA exists, the peak of the AF is located at the $(\tau, f_d) = (0,0)$. The location of the peak value is shifted with a moving TA, and located in the latency-Doppler domain. In (\ref{AF2}), it can be seen that the AF is the combination of multiple spikes. The shape of each spike is determined by the transmit pulse type and precoding algorithm. Owing to the precoding methods, the peak of each spike becomes relative sharper than classical LFM signal, which exhibits a better range resolution. The corresponding zero-latency and zero-Doppler cuts are presented in (\ref{AF3}) and (\ref{AF4}). When $\tau = 0$ in (\ref{AF3}) or $f_d = 0$ in (\ref{AF4}), a larger symbol number $N$ leads to a narrower $f_{d,w}$ or $\tau_w$, indicating a smaller circle outline of each pulse, i.e., the better range and Doppler resolutions. Moreover, a larger $\mu$ leads to a larger sweep bandwidth $B_w$, which can increase the range resolution.
% Here is Figure-6 AF

\subsection{Sum-rate of the proposed DFRC system}
\begin{figure}[t]
\centering
\includegraphics[width = 3.2 in]{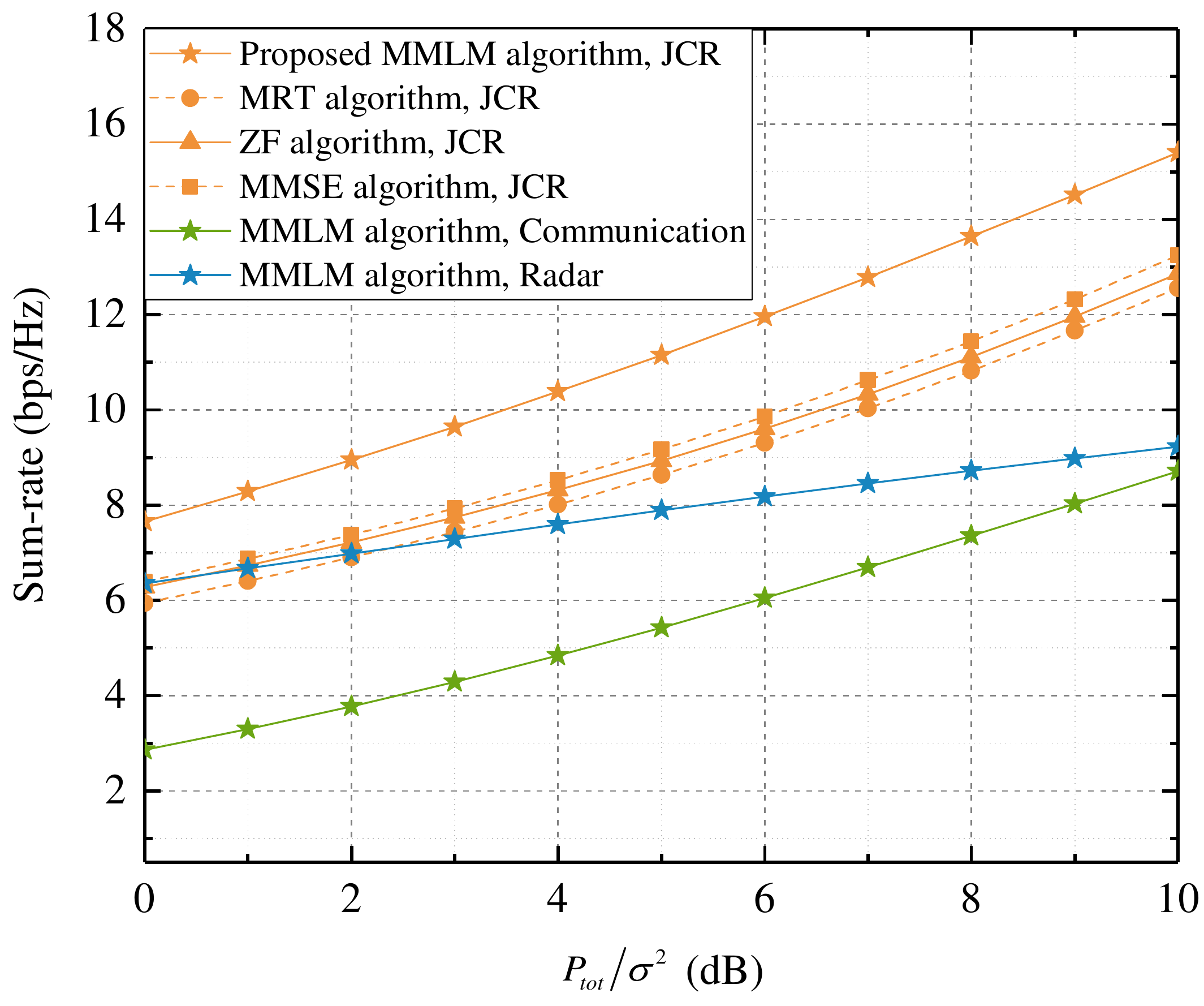}
\caption {An illustration of the sum-rate versus $P_{tot}/{\sigma^2}$ in proposed MMLM, ZF, MMSE, MRT algorithms in DFRC, pure communication and radar systems.} 
\label{comm_rate_pd}
\end{figure}

\begin{figure}[t]
\centering
\includegraphics[width = 3.2 in]{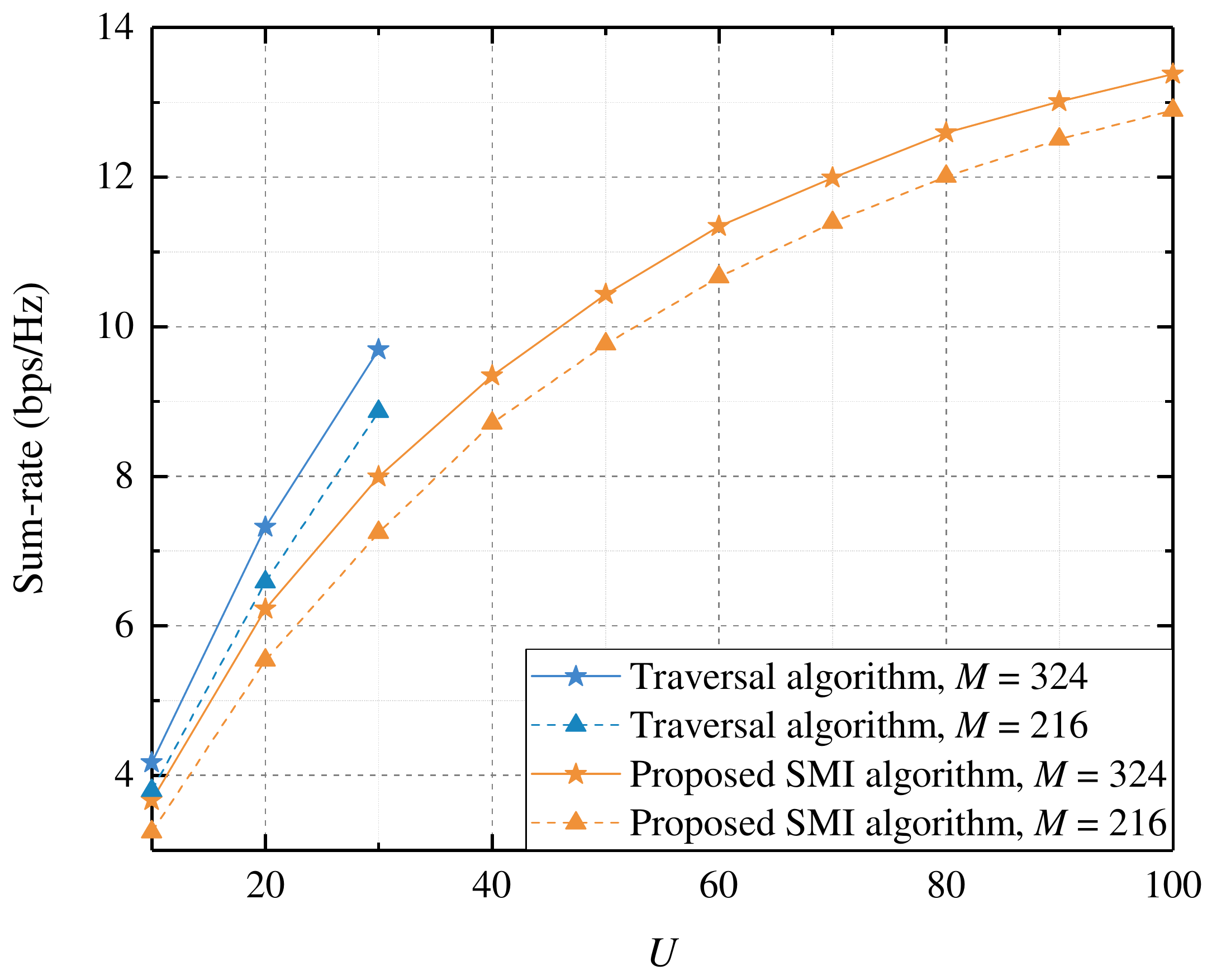}
\caption {An illustration of the sum-rate versus $U$ in traversal and proposed SMI algorithm, where $M = 324$ and $216$.}
\label{UE_selection}
\end{figure}

The sum-rates of the proposed DFRC system are shown in Figs. \ref{comm_rate_pd}-\ref{UE_selection}. Fig. \ref{comm_rate_pd} shows the influence of $\bf W$, with the assumption $K=4$; and Fig. \ref{UE_selection} focuses on the impact of user selection algorithm.

According to the simulation parameters, the bandwidth of the radar system is five times larger than communication system, i.e. $B_w = 5 \, B_g$. To make a fair comparison, we should eliminate the influence of signal bandwidth and consider the sum-rate per bandwidth. It can be seen from Fig. \ref{comm_rate_pd}, the proposed scheme has the best sum-rate performance than the pure communication or radar system. This is reasonable since there is a trade-off between the sum-rate per bandwidth with the detection ability in radar system. Moreover, for the proposed JCR system, our designed $\bf W$ provides the largest sum-rate than other classical beamforming algorithms, i.e., minimum mean square error (MMSE), followed by zero-forcing (ZF) and MRT. For example, when $\frac{P_{tot}}{\sigma^2} = 10$ dB, the sum-rate of our proposed scheme is $15.3$ bps/Hz, which is approximately $2.1$ bps/Hz, $2.5$ bps/Hz and $2.8$ bps/Hz larger than the MMSE, ZF and MRT algorithms.

From Fig. \ref{UE_selection}, it can be seen that the traversal user selection algorithm always provides the larger sum-rate than the proposed SMI algorithm, since it can achieve the optimal solution. Evidently, the complexity of SMI is much lower than the traversal algorithm, especially when $U$ and $K$ are large values. For example, when $U = 30$ and $K = 4$, the complexity of SMI is $120$, while traversal algorithm is $8.1\times 10^5$. Moreover, $M$ shows the impact on the TA detection precision. Let ${\delta_\varphi} = 10^\circ$, we compare the case ${\delta_\theta} = 10^\circ $ and ${\delta_\theta} = 15^\circ $, the corresponding $M = 324$ and $216$. It can be derived that a larger $M$ represents a higher sum-rate since it can enhance elevation or azimuth resolution, indicating the less interferences of each UE.
% Here is Figure-7 Sum-rate W and V

% Here is Figure-8 Sum-rate K

\begin{comment}
% Here is Figure-9 Complexity
\begin{figure}[t]
\centering
\includegraphics[width = 2.6 in]{Complexity.pdf}
\caption {The complexity of the traversal and proposed SMI-MMLM.}
\label{Complexity}
\end{figure}
\end{comment}

\subsection{Trade-off between communication and radar performances}
\begin{figure}[t]
\centering
\includegraphics[width=3.2 in]{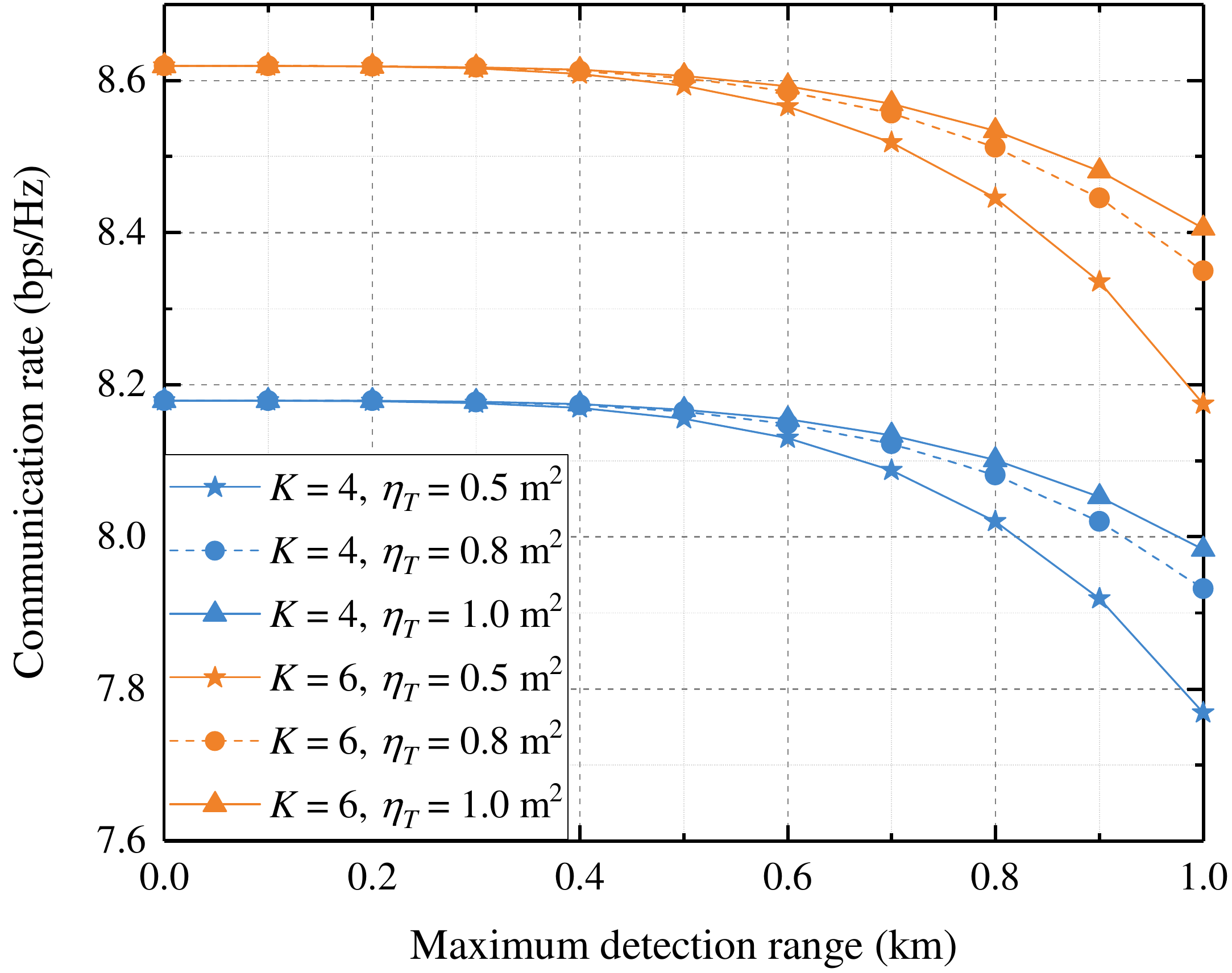}
\caption{Comparison between the communication rate versus maximum detection range in different user number $K$ and RCS, where the chirp rate $\mu = 10^{10}$.}
\label{com_rate_range}
\vspace{-0.2in}
\end{figure}

\begin{comment}
Fig. \ref{com_rate_pd} illustrates the relationship between the achievable communication rate and detection probability of the TA, with various $P_{fa}$ and $\mu$. Simulation results reveal that the communication rate decreases with the increased $P_d$, when the total power $P_{tot}$ is fixed. To achieve a larger $P_d$, more power should be allocated to the radar system, indicating a smaller $P_k$. When $P_d$ is fixed, a lower $P_{fa}$ and/or a larger $\mu$ lead to a higher communication sum-rate, since a lower $P_{fa}$ needs less radar power, and a large $\mu$ can increase the communication bandwidth $B_g$. In the case $P_{fa} = 10^{-6}$ and $\mu = 10^{10}$, when $P_d = 90\%$, the communication rate is $7.5$ bps/Hz, and the sum-rate lose is less than $0.5$ bps/Hz compared to the case $P_d = 10\%$.
% Here is Figure-9 Communication rate Pd
\begin{figure}[t]
\centering
\includegraphics[width = 3.2 in]{com_rate_detection.pdf}
\caption {The comparison between communication rate versus detection probability $P_d$ in chirp rate $\mu = {\rm 10}^{10}$, $5 \times {\rm 10}^{9}$ and false alarm probability $P_{fa} = {\rm 10}^{-12}, {\rm 10}^{-10}, {\rm 10}^{-8}, {\rm 10}^{-6}$.}
\label{com_rate_pd}
\end{figure}
\end{comment}
The relationship between the communication rate and detection range is shown in Fig. \ref{com_rate_range}, where $K = 4,6$ and RCS $\eta_T = 0.5, 0.8, 1.0 \, {\rm m}^2$. When $K$ and $\eta_T$ are constant, the communication rate decreases with the increased maximum detection range, since a larger detection range needs a larger $P_{\rm rad}$, leading to a smaller $P_{\rm com}$. It is remarkable that the influence of RCS can be ignored, when the detection distance is less than a threshold, i.e., $0.3$. In the case $K=4$ and $\eta_T = 0.5$, when the maximum detection range is $0.6$ km, the communication rate is approximately $8.15$ bps/Hz, while the maximum detection range is $0.8$ km, the communication rate is $8.02$ bps/Hz. The trade-off between the maximum detection range and the communication rate is mainly determined by the power allocation, since the total transmit power $P_{tot}$ is a constant value, a better radar performance leads to a smaller communication rate.
% Here is Figure-10 Communication rate range
%, therefore, the designing of $M$ and $N$ should take into consideration both elevation \& azimuth and range \& velocity resolutions

% section-7
\section{Conclusion}\label{section7}
In this paper, we present a joint beamforming and CPM-LFM integrated signal, based on the proposed dual mode framework. The framework includes both dynamic mode and static mode, and the parameter $M$ and $N$ affect scanning precision and detection resolutions respectively. Following, we propose an objective function of sum-rate with constraints. To solve the non-convex issue, we give a sub-optimal joint beamforming and resource allocation algorithm, i.e. SMI-MMLM, which includes successive maximum iteration for user selection; and a maximum minimization Lagrange multiplier algorithm for precoding \& processing design, with water filling algorithm to achieve the optimal power allocation results. Then, we give a detail analysis of the proposed system, including the spectrum characteristic, BER and AF. Simulations results shows that our proposed CPM-LFM signal has better spectrum efficiency and range resolution. Moreover, the proposed SMI algorithm shows lower complexity than the optimal traversal algorithm, and the proposed MMLM algorithm performs larger sum-rate than the classical MRT, ZF and MMSE algorithms.

\begin{comment}
\begin{IEEEbiography}[{\includegraphics[width=1in,height=1.25in,clip,keepaspectratio]{yucao.jpg}}]{Yu Cao}
received the B.S. degrees in Electronic and Information Engineering, from Harbin Institute of Technology (HIT), P. R. China, in 2018. Currently he is a Ph.D student majored in information and communication engineering of HIT. His research interests include joint communication and radar signal processing, MIMO precoding technology and optimization theory.
\end{IEEEbiography}

\begin{IEEEbiography}[{\includegraphics[width=1in,height=1.25in,clip,keepaspectratio]{yucao.jpg}}]{Qi-Yue Yu}
(M'08-SM'17) received her B. Eng., M. Eng., and Ph.D. degrees from Harbin Institute of Technology (HIT), China, in 2004, 2006, and 2010, respectively.

She is currently a full professor at the school of Electronics and Information Engineering, HIT. During Apr. 2007-Mar. 2008, she studied in Adachi Lab, Tohoku University, Japan and was a research assistant of Tohoku University Global COE program. In 2010, she was invited to City University of Hong Kong to research on multi-user MIMO technology. And she was invited to University of Southern Queensland to do researches on the distributed antenna system in Jan.-Mar. 2014. During Sept.2015-Sept.2016, she was a visiting scholar in University of California, Davis. Her research interests include modulation and coding, information theory, multi-access techniques and MIMO for broadband wireless communications.
\end{IEEEbiography}
\end{comment}

% that's all folks
\end{document}